\documentclass[12pt,preprint]{aastex}

\usepackage{epsfig}

\slugcomment{Submitted to the Astrophysical Journal}

\shorttitle{Residence Times of Dust in Protoplanetary Disk Environments}
\shortauthors{F. J. Ciesla}

\begin{document}


\title{Residence Times of Particles in Diffusive Protoplanetary Disk Environments I. Vertical Motions}

\author{F. J. Ciesla\altaffilmark{1}}
\affil{Department of the Geophysical Sciences, The University of Chicago, 5734 South Ellis Avenue, Chicago, IL 60637}
\newpage

\begin{abstract}
The chemical and physical evolution of primitive materials in protoplanetary disks are determined by the types of environments they are exposed to and their residence times within each environment.  Here a method for calculating representative paths of materials in diffusive protoplanetary disks is developed and applied to understanding how the vertical trajectories that particles take impact their overall evolution.  The methods are general enough to be applied to disks with uniform diffusivity, the so-called ``constant-$\alpha$'' cases, and disks with a spatially varying diffusivity, such as expected in ``layered-disks.''  The average long-term dynamical evolution of small particles and gaseous molecules is independent of the specific form of the diffusivity in that they spend comparable fractions of their lifetimes at different heights in the disk.  However, the paths that individual particles and molecules take depend strongly on the form of the diffusivity leading to a different range of behavior of particles in terms of deviations from the mean.  As temperatures, gas densities, chemical abundances, and photon fluxes will vary with height in protoplanetary disks, the different paths taken by primitive materials will lead to differences in their chemical and physical evolution.  Examples of differences in gas phase chemistry and photochemistry are explored here.  The methods outlined here provide a means of integrating particle dynamics with chemical models to understand the formation and evolution of primitive materials in protoplanetary disks over timescales of 10$^{5}$-10$^{6}$ years.

\end{abstract}

\keywords{astrochemistry; comets: general; meteorites, meteors, meteoroids; methods: numerical; protoplanetary disks}

\newpage

\section{Introduction}

Protoplanetary disks are dynamic objects through which mass and angular momentum are transported during the final stages of pre-main sequence evolution of their central stars.  This evolution leads to the radial expansion and thinning of the disk, along with changes to its thermal and density profiles.  Primitive materials contained within the disk thus experience a range of physical environments over their disk lifetimes, being subjected to different levels of alteration which set their final chemical, isotopic, and mineralogical properties.  The properties of planetesimals, and by extension the planets that accrete from them, are determined by the sum of the individual dust grains from which they formed, and thus are dependent on the disk environments to which primitive dust grains were exposed and how long they spent in each environment.

Within our own solar system, chondritic meteorites provide hints of the range of environments that dust grains were exposed to within our own protoplanetary disk, the solar nebula. Refractory inclusions, such as Calcium, Aluminum-rich Inclusions (CAIs), record a very hot ($T>$1300 K) stage of nebular evolution, as they are composed of the first minerals to condense from a gas of solar composition \citep{grossman72,ebel06}.  The matrix within which these refractory inclusions are embedded include pre-solar grains, which were inherited from the parent molecular cloud and have escaped significant processing (e.g. temperatures above a few hundred degrees) all together \citep[e.g.][]{huss04}.  These two cases illustrate the extreme levels of alteration that could occur within a disk: total destruction due to vaporization (and subsequent recondensation) in the case of refractory inclusions, and total avoidance of alteration in the case of pre-solar grains.  Many meteoritic materials fall in between these two extremes, recording different levels of evaporation, condensation, gas-solid reactions, and internal restructuring such as annealing and solid-state diffusion.
Decades worth of research has been dedicated to understanding what environments would have allowed these different reactions to occur in order to explain the observed properties of primitive materials, with much of it summarized and reviewed recently in \citet{laurettamcsween06}.
While this research has provided insights into the conditions that would have existed within the solar nebula and other protoplanetary disks, a critical issue to understand is how the dynamic evolution of a protoplanetary disk could expose dust grains to these different environments while still allowing them to become mixed on fine (sub-centimeter) scales as seen in chondritic meteorites.  

 One of the leading candidates for explaining this mixing and transport is turbulence within the solar nebula.  Turbulence  has long been invoked as the source of the viscosity needed to drive mass and angular momentum transport within protoplanetary disks \citep{ss73,lbp74,rudpol91}.  Many properties of chondritic meteorites can be understood in the context of a turbulent solar nebula \citep{cuzzi05,cieslacuzzi06} and the delivery of high-temperature materials to the comet formation region has also been explained in this setting \citep{gail01,bockelee02,ciesla07b,ciesla09}.  In fact, planetesimal formation itself may rely on turbulence within the disk \citep{cuzzi01,cuzzi08,johansen07}.  Thus, while the level of turbulence in protoplanetary disks is still the subject of ongoing study \citep{gammie96,stone01,flemstone03,oishi07}, the overall impact it has on the evolution of primitive materials must be understood.

The detailed effects that turbulence would have on the chemical evolution of primitive materials found in meteorites and comets have not been quantified.  That is, most models for the chemical evolution of primitive materials assume some set of ambient conditions (e.g. pressure, temperature, composition of the gas and solids) and then calculate what chemical reactions should take place.  In some cases, these calculations predict what products should form if thermodynamic equilibrium is reached--that is given an infinite amount of time for reactions to take place \citep{grossman72,ebel00,ebel06}.  In other cases, kinetic rates are known or estimated, and the chemical evolution of the solids and gas can be calculated within a given environment \citep[e.g.][]{fegley00}.  These kinetic calculations have been carried out for a variety of situations, as recently reviewed by \citet{cieslacharnley06} and \citet{bergin07}.  In both types of study, chemical models have demonstrated that many of the materials seen in chondritic meteorites and comets can be produced within environments that plausibly existed within the solar nebula.
Unfortunately, the assumptions in these various chemical models are not always consistent with the evolution of materials in a dynamic protoplanetary disk environment:  while they assume that the conditions in which the chemistry takes place is constant in their simulations, turbulence would lead to materials diffusing around a protoplanetary disk, being exposed to a wide array of environments (different pressures, temperatures, and radiation fluxes).  Within these different environments the thermodynamically stable materials may change, or the rates of kinetic reactions may vary.  Thus the treatments of these chemical models, and thus their specific predictions, may differ from the real evolution that occurs within a protoplanetary disk.

This paper presents a model that can be applied to quantifying the effects that turbulent motions of gas and dust in a protoplanetary disk have on the long-term chemical evolution of primitive materials.  Specifically, the trajectories of small
particles and gas in a protoplanetary disk are calculated using a Monte Carlo model.  Here the focus is on the vertical motions of particles within the disk, where the pressure, temperature, and radiation flux may all vary, sometimes significantly, with height above the disk midplane.  By calculating the motions of individual particles, we can determine the types of paths they take through a disk and collect statistics on what environments these particles are exposed to and for how long.  This information can then be used to determine what kinds of chemistry would occur, and what variations should be seen when looking at primitive materials.  Radial motions will be considered in future work.

This paper is organized as follows.  The Monte Carlo model is described in $\S$2.  The model is developed  for both the case where the level of turbulence is constant everywhere and the case where the level of turbulence (quantified by the $\alpha$ parameter) varies with height, possibly due to the vertical dependence on the ionization fraction of the gas and thus the ability of the Magnetorotational Instability (MRI) to develop \citep[e.g.][]{bh91,gammie96,glassgold97,stone01,flemstone03,oishi07}.   In $\S$3 ,results of these models are presented for various cases, with the focus being on specific applications relevant to the chemical processing of primitive materials in protoplanetary disks.  The results are then summarized and implications discussed in $\S$4.

\section{Model}

In the standard picture of a turbulent protoplanetary disk, the random kicks from turbulent eddies allow materials to be redistributed with an effective diffusivity, $D$.  In most models, the diffusivity is taken to be comparable to the disk viscosity, though there have been studies deriving more specific relations between these two quantities \citep[e.g.][]{jkm06}.  For the purposes of this paper, we will make the first order assumption that the diffusivity is given by $D= \nu =\alpha c_{s} H$, where $c_{s}$ is the local speed of sound, $H$ is the local gas scale-height, and $\alpha$ is a parameter $<$1 which quantifies the level of turbulence in the disk and may or may not be comparable to the $\alpha_{visc}$ that describes the diffusion of angular momentum in the disk.  While this basic formalism is adopted, the techniques and methods developed here are general enough such that other relationships can easily be accommodated.  Note the focus is on small particles here such that the Stokes number, $St$, defined as the ratio of particle stopping time ($t_{s}$=$\rho a$/$\rho_{g} c_{s}$, with $\rho$ the material density, $a$ the particle radius, and $\rho_{g}$ the local gas density) to turnover time of the largest eddies ($t_{eddy}$, assumed to be comparable to the Keplerian orbital period, $\sim$1/$\Omega$) satisfies $St << 1$.  Such a treatment makes corrections to the diffusivity for particle inertia negligible \citep{dubrulle95,cuzziweid06,yl07}.  However the methods here could be easily extended to account for variations where the inertia of bodies significantly  impacts their diffusivity.

\subsection{Monte Carlo Model of Material Redistribution}

\subsubsection{General Advection-Diffusion Equation}

The flux of a diffusing species is given by Fick's law:
\begin{equation}
F=-D \frac{\partial f}{\partial z}
\end{equation}
where $f$ describes the abundance of a particular species at a given coordinate, $z$, and $D$ is the diffusivity of materials \citep{fick1855}.  Given that flux, the diffusion-advection equation is written in flux conservative form:
\begin{equation}
\frac{\partial f}{\partial t} = \frac{\partial}{\partial z} \left( D \frac{\partial f}{\partial z} \right)
- \frac{\partial}{\partial z} \left( v f \right)
\end{equation}
where the change in abundance with time is equal to the differences in fluxes due to diffusion (first term on right) and advective flow (second term on right).

Monte Carlo, or particle-tracking, models solve the diffusion-advection equation by calculating the positions of particles subjected to the different transport processes over time.  In the absence of diffusion, or with $D=0$, the position of a particle after some small time interval, $\delta t$, is simply given by:
\begin{equation}
z_{i}=z_{i-1}+v \delta t
\end{equation}
where $z_{i-1}$ is the position of the particle at the beginning of the timestep and $z_{i}$ is the position at the end of it.  In such a scenario, all particles starting at $z=z_{0}$ in a given simulation would be found at the same $z=z_{f}$ point at the end of the simulation.

In cases with non-zero $D$, such as within a turbulent fluid, the random motions associated with entrainment in eddies leads to random displacements of the species of interest.  As a result, given enough time (longer than the overturn time of the largest turbulent eddy) particles starting at the same $z=z_{0}$ follow unique, independent paths.  Monte Carlo models account for this by adding random displacements on top of the advective motions.  For the case of a non-zero $D$ the new position for a particle after some time, $\delta t$, would be given by:
\begin{equation}
z_{i}=z_{i-1}+v \delta t + R \left( \frac{2}{r} D \delta t \right)^{\frac{1}{2}}
\end{equation}
where $R$ is a value from a random distribution and $r$ is the variance (square of the standard deviation) of that random distribution.  For the modeling done here, $R$ is taken as a random number from a uniform distribution with a value between -1 and 1, and thus $r$=1/3 \citep{visser97}.  The third term on the right represents the random displacement that a particle experiences over the time interval $\delta t$ due to turbulent eddies (or driving mechanism behind the diffusion).  The form of this term comes from the fact that the variance of materials due to diffusion goes as $d<z^{2}>$/$dt$=$2D$ \citep{visser97}.

An example of a Monte Carlo solution to equation (2) is shown in Figures 1 and 2.  Here, the paths of 10$^{4}$ particles starting at $z_{0}=$0 were calculated for 100 seconds.  The diffusivity was assumed to have a value of 10 cm$^{2}$/s and the advective velocity was assumed to be $v$=10 cm/s.  Equation (4) was used to calculate the position of each particle throughout the simulation.  Random numbers were taken using the \emph{ran2} routine of \citet{press}, and a timestep of $\delta t$=0.01 s was used.  Figure 1 shows the distributions of these particles after 10, 50, and 100 seconds, while Figure 2 shows the temporal evolution of the average position of the particles and the standard deviation (square root of the variance) of their distribution.  In both cases, the numerical solutions are compared to the expected (analytic) values, with good agreement being seen.

\subsubsection{Spatially Varying Diffusivity}

As demonstrated above, provided a statistically significant number of particles is considered, Monte Carlo models can be used to solve for the dynamical evolution of a swarm of particles subjected to diffusive-advective motions.  This is because by using a large number of particles, the Monte Carlo methods satisfy the  Fokker-Planck equation (FPE):
\begin{equation}
\frac{\partial P \left( z, t \right) }{\partial t} = \frac{\partial^{2}}{\partial z^{2}} \left[ B\left( z \right) P \left( z,t \right) \right] - \frac{\partial}{\partial z} \left[ A\left( z \right) P\left( z, t \right) \right]
\end{equation}
 \citep{kinzelbach90,bodin03,delay05}.  The FPE describes how $P \left( z, t \right)$, the probability density for a particle to be located at $z$ at time $t$, evolves with time when the particles are subjected to some mean drift rate,  $A \left(z \right)$, and some dispersion process, $B \left( z \right)$.  Similarities between the two equations can be readily seen if we  substitute:
 \begin{equation}
 f \left(z, t \right) = P \left(z, t \right)
 \end{equation}
 \begin{equation}
v =  A\left(z \right)
 \end{equation}
 \begin{equation}
 D = B \left(z \right)
 \end{equation}
 yielding:
 \begin{equation}
 \frac{\partial f}{\partial t} = \frac{\partial^{2}}{\partial z^{2}} \left( D f \right) - \frac{\partial}{\partial z} \left( vf \right)
\end{equation}
 
 Note however, that the diffusion-advection equation is equivalent to the FPE only in the case that the diffusivity is constant.  If $D$=$D \left( z \right)$, then expansion of the first term on the right  of Eq. (9) gives:
\begin{equation}
\frac{\partial f}{\partial t} = \frac{\partial}{\partial z} \left[ D \left( z \right) \frac{\partial f}{\partial z} \right] + 
\frac{\partial}{\partial z} \left[f \frac{\partial D \left( z \right)}{\partial z} \right]
- \frac{\partial}{\partial z} \left[ v \left(z \right) f \right]
\end{equation}
which provides an additional term on top of the advection-diffusion equation.  Note that the velocity is also allowed to vary with location, however this does not change the form of the equation.  The additional term arising from expansion of the diffusion term in the FPE has the form of advective motions.  Re-writing the advection-diffusion equation as:
\begin{equation}
\frac{\partial f}{\partial t} = \frac{\partial^{2}}{\partial z^{2}} \left[ D\left( z \right) f \right] 
- \frac{\partial}{\partial z} \left[ v_{eff} \left(z \right) f \right]
\end{equation}
where
\begin{equation}
v_{eff}=v \left(z \right) + v_{FP} \left( z \right)
\end{equation}
with
\begin{equation}
v_{FP}=\frac{\partial D \left( z \right)}{\partial z}
\end{equation}
allows us to account for the spatial variations of the diffusion coefficient, while still being in agreement with the FPE.   Equation (11) is thus similar to Equation (2), except the advective component is modified so that materials drift in the direction of increasing diffusivity.  A different justification  of this additional velocity component which considers  the moments of the diffusion equation is presented in the Appendix.

This additional advective component must also be added to the motions of particles in the Monte Carlo model.  If the diffusivity has a local gradient given by:
\begin{equation}
D'=\frac{\partial D}{\partial z}
\end{equation}
Equation (4) is then re-written as:
\begin{equation}
z_{i} = z_{i-1} +  v_{eff} \delta t + R \left[ \frac{2}{r} D \left( z' \right) \delta t \right]^{\frac{1}{2}}
\end{equation}
where  
\begin{equation}
z'=z_{i-1} + \frac{1}{2} D' \left( z_{i-1} \right) \delta t
\end{equation}
 The third term on the right of Equation (15), as in Equation (4), represents the random motions associated with the diffusive process.  Note here that the random step is not evaluated at the current location of the particle, but is offset slightly in the direction of increasing diffusivity.  Many authors have simply evaluated the magnitude of the random displacement at the previous location of the particle, $z_{i-1}$, however, \citet{labolle00} and \citet{hassan03} found that by offsetting the evaluation of this displacement, numerical models provide more accurate solutions in the vicinity of strong gradients in diffusivity.  Such a correction is generally minor in many cases, and using $D\left(z_{i-1}\right)$ does not result in significant differences in numerical models presented here.   (A further derivation justifying using $D\left(z'\right)$ is found in the moments treatment presented in the Appendix.) Note that when $D'$=0, we recover Equation 4.  Thus the main difference in the case of spatially varying diffusivity is that particles drift in the direction of increasing diffusivity, and the random displacements increase in magnitude as they enter this more diffusive region.

\subsection{Monte Carlo Simulations in a Protoplanetary Disk }

Having established the Monte Carlo methods needed to simulate the diffusion and advection of particles, it is now time to apply these methods to the case of motions in a protoplanetary disk.  The  vertical motions of materials in a turbulent disk are described by  \citep{dubrulle95,fromang06}:
\begin{equation}
\frac{\partial \rho_{i}}{\partial t} =  \frac{\partial}{\partial z} \left( \rho_{g} D \frac{\partial C_{i}}{\partial z} \right)  
-\frac{\partial}{\partial z}\left( \rho_{i}v_{z} \right)
\end{equation}
where $C_{i}$ is the concentration of interest, or the ratio of the density of the material of interest, 
$\rho_{i}$, to the local gas density, $\rho_{g}$ (that is, $C_{i}$=$\rho_{i}$/$\rho_{g}$). The vertical velocity, $v_{z}$, would be that due to gravitational settling. Equation (17) differs slightly
from the diffusion equations from above as the diffusive flux is determined by the concentration of the species within the gas rather than the absolute abundance.  (In the case that $\rho_{g}$ is constant with $z$ and $t$, then Equation (17) is equivalent to Equation (2)).  It should be noted that here only the motions of particles are considered, and any growth via collisional aggregation is ignored.

In order to apply the Monte Carlo model to this particular case, we must re-write Equation (17) by expanding the first term on the right hand side.  This gives:
\begin{equation}
\frac{\partial \rho_{i}}{\partial t} =  \frac{\partial}{\partial z} \left( D \frac{\partial \rho_{i}}{\partial z} \right)  
- \frac{\partial}{\partial z} \left( \frac{D}{\rho_{g}} \frac{\partial \rho_{g}}{\partial z} \rho_{i} \right)
-\frac{\partial}{\partial z}\left( \rho_{i}v_{z} \right)
\end{equation}
Assuming the disk is in hydrostatic equilibrium such that 
$\rho_{g}$($z$)=$\rho_{g,0} e^{-\frac{z^{2}}{2H^{2}}}$ and that this profile does not change with time gives:
\begin{equation}
\frac{\partial \rho_{i}}{\partial t} = \frac{\partial}{\partial z} \left( D \frac{\partial \rho_{i}}{\partial z} \right) +\frac{\partial}{\partial z} \left( D \frac{z}{H^{2}} \rho_{i} \right) - \frac{\partial}{\partial z} \left( \rho_{i}v_{z} \right)
\end{equation}

The first and third terms on the right in Equation (19) then are analogous to the diffusive and advective terms  in Equation 2.  The second term on the right also has the form of an advective component with velocity:
\begin{equation}
v_{gas} = -D \frac{z}{H^{2}}
\end{equation}
This term arises from the fact that the medium in which the species is diffusing, the nebular gas, varies in density with height, and it is the ratio of the species density to the gas density that determines the mass flux due to diffusion.  While the source of the effective velocity is not necessarily intuitive, it becomes apparent when we consider the vertical diffusion of a gaseous component, such as water vapor.  By considering a gaseous species, we ensure that $v_{z}=0$ everywhere, as the gas is assumed to be in hydrostatic equilibrium, and thus only diffusion would be responsible for transport.  Given enough time, water vapor should become uniformly mixed, giving a constant value of $C_{H_{2}O}$ with height. That is, the final state should be a gaussian distribution of water molecules centered on the midplane and having a standard deviation, $\sigma=H$. Carrying out a Monte Carlo simulation without the effective velocity, we would simply have Equation (4) with $v=0$ everywhere.  This would result in a standard deviation that grew with time as $\sigma$=$\sqrt{2 D t}$, meaning that the distribution of water molecules would expand indefinitely.  The effective velocity thus limits the ability of materials to expand  above and below the disk midplane as materials are instead pulled back toward the disk midplane at a rate that increases with height, ensuring in this case that the water molecules follow the same gaussian distribution as the nebular gas.

As an example, Figure 3 shows the standard deviation of 20,000 gas molecules undergoing vertical diffusion within a protoplanetary disk as a function of time.  The motions of these particles were calculated using:
\begin{equation}
z_{i} = z_{i-1} +  v_{gas} \delta t + R \left[ \frac{2}{r} D  \delta t \right]^{\frac{1}{2}}
\end{equation}
for the case of uniform $D$.
Again, note there is no additional vertical velocity due to gravitational settling because of hydrostatic equilibrium.  In this particular case, we consider a radial location of 1 AU, a surface density of gas of $\Sigma$=2000 g/cm$^{2}$, a uniform vertical temperature of $T$=280 K, and values of $\alpha$=10$^{-4}$ and 10$^{-2}$.  A timestep off $\delta t$=0.1 yr was used, which is comparable to the correlation time expected for turbulence in protoplanetary disks \citep[$\sim$ 0.15 orbital periods][]{fromang06}, though runs with factors of 5 differences in timestep yielded little change in the results.  All molecules were released at the midplane ($z=0$) at the beginning of the simulation.  Initially, the standard deviation grows as particles diffuse away from the midplane, but with time, it levels out and follows $\sigma \sim H$ for the rest of the simulation.  Different values of $\alpha$ have similar behavior, with the timescale for  equilibrium ($\sigma$=$H$) being reached being inversely proportional to the choice of the turbulence parameter (that is, for $\alpha$=10$^{-2}$, $t_{eq} \sim$ 20 years and for $\alpha$=10$^{-4}$, $t_{eq} \sim$ 2000 years).

Figure 4 shows a similar calculation as in Figure 3, but now considering the movement of centimeter-diameter particles of unit density.  These solids experience gravitational settling as the vertical component of gravity from the central star pulls them toward the midplane while their motions are retarded by drag as they attempt to move through the gaseous fluid.  The gravitational settling velocities here were given by the terminal velocities of the particles \citep{cuzziweid06}:
\begin{equation}
v_{grav}=-t_{s} \Omega^{2} z
\end{equation}
where $t_{s}$ is again the particle stopping time.  Equation (21) was used, but with $v_{gas}$ replaced with $v_{eff}$=$v_{gas}+v_{grav}$. 

 Because of the gravitational settling of the particles, these objects do not extend to as great a height as the gas in the disk, and instead are found to follow gaussian distribution with an effective scale height, $H_{d}$=0.9$H$ and $H_{d}$=0.35$H$ (found by averaging the standard deviation over times 5000 years $< t <$ 10,000 years) for the cases of $\alpha$=10$^{-2}$ and 10$^{-4}$ respectively.  Note, these values differ slightly from the thicknesses estimated by \citet{dubrulle95}:
 \begin{equation}
 H_{d}^{Dubrulle} =  \sqrt{\frac{\alpha}{St+\alpha}} H
 \end{equation}
Taking $St \sim$6x10$^{-4}$ this predicts values of  $\sim H$ and $\sim$0.4$H$ for $\alpha$=10$^{-2}$ and 10$^{-4}$ respectively.  \citet{yl07} offered a slightly different form for $H_{d}$ (their Eq 28)  which for the case considered here gives:
 \begin{equation}
 H_{d}^{YL} \sim \sqrt{\left(\frac{\alpha}{\alpha+St} \right) \left( \frac{1 + St}{1+2St}\right)} H
 \end{equation}
 which also gives values of $\sim H$ and $\sim$0.4$H$ respectively.

The discrepancies between the predicted values and those found in the numerical simulations are small and can be explained by the approximations made in the analytic treatments, such as  $\rho_{g}$ being constant with height.  As discussed above, the fact that $\rho_{g}$ decreases with increasing distance from the disk midplane affects the dynamics of the particles.    The importance of these effects depends on the size of the particles, the strength of turbulence ($\alpha$), and the density gradient in the disk.


\subsection{Spatially Varying $\alpha$}

One of the primary candidates for generating turbulence within a protoplanetary disk is the MRI \citep{bh91}.  The MRI operates in a region of a protoplanetary disk that is ionized to a sufficient level that gaseous fluid elements become coupled to one another via magnetic field lines.  This criteria is generally met at high temperatures ($>$1000 K) within the disk, when collisions between gas molecules can lead to the ionization of alkali elements, and in regions where cosmic rays and X-rays are absorbed, leading to the ionization of H$_{2}$ \citep[e.g.][]{gammie96,glassgold97}.  Cosmic rays and X-rays are absorbed after passing through a finite column of H$_{2}$ gas, $\Sigma_{c}$, meaning that the entire thickness of a disk will be ionized to the level needed to sustain the MRI only where the surface density of the disk is less than the critical value ($\Sigma<\Sigma_{c}$).  The exact value of $\Sigma_{c}$ is uncertain, as if X-rays are the only source of ionizing radiation $\Sigma_{c} \sim$10 g/cm$^{2}$ \citep{glassgold97}, but if cosmic rays are important, $\Sigma_{c}$ would be closer to 100 g/cm$^{2}$ \citep{u81,gammie96}.   This suggests only the sparse, outer regions of the disk will be fully ionized and have the MRI active at all heights. In the inner regions of the disk, where $\Sigma > \Sigma_{c}$, only the surface layers of the disk become ionized to the level that MRI is able to fully operate.  Below the surface, the gas may still get stirred up by eddies and waves propagating downward from the active layer above, but the turbulence would be much weaker (lower effective $\alpha$) \citep[e.g.][]{stone01,flemstone03,oishi07}.

A change in $\alpha$ in such disks  corresponds to differences in the viscosity which controls the diffusion of angular momentum.  Again, while the viscosity does not necessarily equate to the diffusivity of materials within the disk, we adopt this as a first-order approximation (particularly as the exact value of $\alpha_{visc}$ in MRI turbulence is uncertain).  Explicit relationships between the viscosity and the diffusivity of disk material, such as presented in \citet{jkm06} can easily be accommodated by the methods used here.

To examine the effects of vertical variations in diffusivity as might arise in a layered disk, we define an expression for $\alpha$ as a function of location: 
\begin{equation}
\alpha \left( z \right)= 10^{-4} + 9.9 \times 10^{-3} e^{-\frac{\Delta \Sigma}{\Sigma_{c}}} 
\end{equation}
where $\Delta \Sigma$ is the surface density of gas between the surface of the disk ($z= \infty$) and the current location, $z_{i}$, and is given by $\Delta \Sigma = \frac{1}{2} \Sigma - \Sigma' \left( z \right)$, with:
\begin{equation}
\Sigma'  \left( z \right) = \int_{0}^{| z |} \rho_{g} dz 
\end{equation}
Figure 5 shows how $\alpha$ varies with height in the case of $\Sigma$=2000 g/cm$^{2}$ and $T$=280 K at 1 AU for values of $\Sigma_{c}$=10 and 100 g/cm$^{2}$.  This form of $\alpha$ is chosen to qualitatively match the vertical structure of stresses in layered disks \citep[e.g.][]{oishi07} where $\alpha$ has values of $\sim$0.01 in the MRI-active regions, but decreases by orders of magnitude in the non-MRI active region. 

Figures 6 and 7  show the evolution of the swarm of gas molecules and a swarm of centimeter-sized particles released at the disk midplane in the two diffusive environments described in Figure 5.  The positions of the particles were calculated using:
\begin{equation}
z_{i} = z_{i-1} +  v_{eff} \delta t + R \left[ \frac{2}{r} D \left( z' \right) \delta t \right]^{\frac{1}{2}}
\end{equation}
where
\begin{equation}
v_{eff} = v_{grav} + v_{gas} + v_{FP} 
\end{equation}
Again, for gas molecules, the expectation is that the standard deviation of their distributions should approach $\sigma$=$H$, given enough time.  This is found to be the case, with the diffusive environment with $\Sigma_{c}$=100 g/cm$^{2}$ reaching this value on shorter timescales than the case of $\Sigma_{c}$=10 g/cm$^{2}$.  Intuitively this makes sense, as the value of the diffusivity is larger throughout much of the height of the disk in the former case than in the latter, resulting in more rapid redistribution of the materials.  This effect is also seen in Figure 7, where the $\Sigma_{c}$=100 g/cm$^{2}$ case allows the centimeter-sized particles to get lofted to higher altitudes, and thus be distributed to greater heights than the $\Sigma_{c}$=10 g/cm$^{2}$.  In these cases, the thickness of the particle layers are $H_{d} \sim$0.47$H$ and 0.35$H$ respectively.  The case of $\Sigma_{c}=$10 g/cm$^{2}$ is very similar to the $\alpha$=10$^{-4}$ case, as the value of $\alpha$ from the disk midplane to nearly $z=\pm 2H$ is $\sim$10$^{-4}$.

\section{Applications}

In the previous section, we developed a Monte Carlo model that calculates representative paths of individual particles within a turbulent protoplanetary disk.  Such information is necessary given that the conditions within a protoplanetary disk (e.g. gas density, temperature, and the irradiation flux) will vary with height above the disk midplane.  Thus the environments to which dust grains are exposed, and their physical and chemical evolution, will depend on their movements through the disk.  Here we apply the Monte Carlo model to understand how long particles would be exposed to different environments within a vertical column of gas in a protoplaneary disk, and what impact this has on the evolution of primitive materials.

\subsection{Residence Times in Different Vertical Environments}

Figure 8 shows histograms for the amount of time the average micron-diameter particle of unit density (averaged over 20,000 particle paths) spent at various vertical heights within a vertical slice disk over a 10$^{5}$ year period for the four cases of diffusivity considered in the previous section ($\alpha=10^{-4}$, $\alpha=10^{-2}$, and $\alpha$=$\alpha \left( z \right)$ as in Eq. (25) with $\Sigma_{c}=$10 and 100 g/cm$^{2}$).  The particles again were assumed to be released at the disk midplane at $r$=1 AU, where $\Sigma$=2000 g/cm$^{2}$ and $T$=280 K.  The 10$^{5}$ year time period considered here is much longer than the time it has taken any of these cases to reach the steady-state distributions described in the previous section (the micron-sized particles, with Stokes numbers of $St \sim$6$\times$10$^{-8}$ at the disk midplane, are very well coupled to the gas and thus have effective $H_{d} \sim H$).  In each case, the time distribution at each height is well described by a gaussian distribution with standard deviation $\sigma \sim H$.    For the particular cases considered here, these results are expected, as the average behavior over a long time period of a particle will reflect the spatial distribution of all particles.  These results are independent of the choice or functional form of $\alpha$.

The primary difference between the different histograms shown here is that the time spent at high altitudes is greater when $\alpha$ reaches larger values there.  That is, with $\alpha$=10$^{-4}$ throughout, the random walk steps are not sufficient to overcome gravitational settling at extreme heights, and thus particles in these simulations do not spend much time at altitudes with $|z|>$4$H$. The resulting distribution is essentially compressed on its extreme edges.  Gaseous molecules, which do not experience gravitational settling,  are more easily able to reach these higher altitudes even in cases of low diffusivity.

While the average, \emph{long-term} behavior of the particles in terms of the heights that they reach and the time spent there does not vary strongly with the choice of $\alpha$ in these models, the manner in which individual particles move through the disk and are exposed to these different environments is very sensitive to the form of $\alpha$.  Figure 9 shows representative paths of single particles from each of the $\alpha$ formalisms discussed thus far.  In the case of $\alpha$=10$^{-4}$, the particle slowly meanders through different heights of the disk, in direct contrast to the $\alpha$=10$^{-2}$ case where the particle experiences large-scale excursions on short timescales.  This would imply that if there are different physical environments at different heights in the disk, an issue that is discussed in detail below, any time the particles in the $\alpha$=10$^{-4}$ disk would be introduced into a given environment, they would spend a longer period of time there than in the higher $\alpha$ case.  On the other hand, particles in the $\alpha$=10$^{-2}$ case would be exposed to a greater number of environments on short timescales.  While the long term behaviors of the particles are similar, the short term evolution, types of conditions they are exposed to, and for how long, could be quite different.  

Different behaviors for the dynamics of the individual particles are also noticeable in the cases of spatially varying $\alpha$.  As would be expected, the midplane region of the disk, where $\alpha$ is near its minimum value, results in particle motions that are similar to the $\alpha$=10$^{-4}$ case described above--that is, particles experience small-scale random displacements, spending relatively long periods of time in this region.  This is in contrast to those times when the particles reside well above the midplane where the diffusivity is higher--there the particles can experience relatively large-scale displacements, lofting them to many scale heights above the disk, and allowing them to be exposed to a greater range of heights on shorter timescales.   The implications for these variations in vertical trajectories on the chemical evolution of materials are explored below.

\subsection{Gas-Solid Chemical Reactions}

As solids are surrounded by gas within a protoplanetary disk, those solids will evolve chemically in a manner that is determined by the ambient pressure, temperature and elemental abundances.  Possible reactions that could occur would be the condensation of gaseous species, evaporation of the solid materials, and corrosion or exchange between the gas and solid particles.  Variations of these different types of reactions have been studied for decades, resulting in a collection of work that is too extensive to review here.  Much of the work relevant to the understanding of primitive materials found in meteorites is discussed in \citet{laurettamcsween06}.  Here, we briefly explore how gas-solid reactions would be affected by the vertical diffusion of grains in a protoplanetary disk.

The first way in which the vertical motions of materials in a disk would impact the predicted chemical evolution is by changing the stability fields of different species and thus the expected chemical reactions that would take place.  For example, equilibrium models predict what materials are expected to form at a given pressure, temperature, and collection of elemental abundances \citep[e.g.][]{grossman72,lodders03}.  As one of these parameters changes, so do the predicted equilibrium chemical products.  In the simple case outlined here of a vertically isothermal disk, the pressure will drop by a factor of $e^{-\frac{z^{2}}{2H^{2}}}$ with height, meaning pressures of $\sim$0.6x, $\sim$0.1x, $\sim$0.01x, and $\sim$3x10$^{-4}$x the midplane pressure at $z=$1$H$, 2$H$, 3$H$, and 4$H$ respectively.  In the case of condensation reactions, the temperature at which condensation occurs generally decreases with decreasing pressure.  As an example, the condensation temperature of water in a gas of solar composition at 10$^{-4}$ bars is $\sim$180 K based on the parameterization given by \citet{bauer97} and confirmed by \citet{lodders03}. Taking the pressure fall off as above, water condensation would occur at temperatures of $\sim$175, $\sim$170, $\sim$160, and $\sim$145 K at these different altitudes \citep{bauer97}.   Thus, scenarios could exist where a given compound condenses out at as a solid at the disk midplane, but vaporizes at higher altitudes.  Thus, that species could undergo many cycles of condensation and evaporation as it moves throughout the column of the disk, which offers the opportunity for isotopic fractionation or other chemical effects \citep{davisrichter05}.  Examples of pressure cycling for particles are shown in Figure 10, which takes the trajectories shown in Figure 9 and plots them in terms of the temporal evolution of the ambient pressure surrounding those particles over the course of their dynamical evolution.

The level to which these pressure cyclings affect the chemical evolution of primitive materials, however, depend on the rates at which the different gas-solid reactions would occur.  This information is best inferred from laboratory experiments, but for the purposes of this discussion, a more general approach suffices.  The basic principles behind gas-solid reaction rates were reviewed by \citet{fegley00} in discussing Simple Collision Theory (SCT) as it applied to the formation of solar nebula materials.  The main idea behind SCT is that given a particle of a known size embedded in a gas of known pressure, temperature and composition, the rate of collisions between a gas phase reactant and the particle can be calculated.  As the velocities of gas molecules would follow a Maxwell-Boltzmann distribution, the distribution of energies for these collisions would also be known.  The chemical reaction would proceed when a collision took place with an energy that exceeded the activation energy for that particular reaction.  Given this information, one could then estimate the amount of time it would take for a reaction to go to completion--that is, the amount of time it would take for it to experience enough high energy collisions such that all of the reactants are exhausted.  

SCT has been applied to look at the formation timescales of such things as troilite, magnetite, and hydrated silicates in either the solar nebula or in planetary subnebulae \citep{lauretta96,fegley00}.  These calculations considered dust grains sitting in an unchanging gaseous environment (constant temperature and pressure) until a reaction was completed.  They found that the timescale for these reactions could range from a few hundred years to billions of years under conditions expected around the midplane of the solar nebula.  Those that had chemical lifetimes less than the lifetime of the solar nebula were considered possible nebula products.  Given that particles would move through different environments in a turbulent disk, including those with different ambient pressures, the collision rates between gas molecules and the solids would vary with time, meaning such motions must be considered when calculating chemical lifetimes.

One can estimate the number of collisions between a particle of radius, $a$ and the H$_{2}$ gas  which would surround it in a protoplanetary disk over a time interval, $\delta t$, as:
\begin{equation}
C_{H_{2}} = 4 \pi a^{2} \left( \frac{ 8 k T}{\pi m_{H_{2}}} \right)^{\frac{1}{2}} n_{H_{2}} \delta t
\end{equation}
where the term in parentheses represents the mean velocity of a H$_{2}$ molecule, with $m_{H_{2}}$ being its mass and $n_{H_{2}}$ the volume number density.  Note this expression can be generalized to collisions for any gaseous species $Y$ by:
\begin{equation}
C_{Y} = C_{H_{2}} \frac{2 A_{Y}}{A_{H}} \left( \frac{m_{H_{2}}}{m_{Y}} \right)^{\frac{1}{2}}
\end{equation}
where $A_{Y}$ and $A_{H}$ are the cosmic abundances of $Y$ and hydrogen atoms respectively, and $m_{Y}$ is the mass of the gaseous species containing $Y$.  Using this formula, we have calculated the number of collisions dust grains experience within a turbulent environment using the Monte Carlo model developed here.  

We again assume particles of uniform size being released from the disk midplane at $t=0$ at 1 AU around a solar mass star, with $\Sigma$=2000 g/cm$^{2}$ and $T$=280 K at all heights.  This gives a number density of hydrogen molecules at the disk midplane of $n_{H_{2}}$=3.48x10$^{14}$ cm$^{-3}$.  Under these conditions, the number of collisions a particle would experience at the disk midplane would be given by:
\begin{equation}
C_{H_{2}} = 2.205 \times 10^{28} \left( \frac{a} {1 ~\mathrm{cm}} \right)^{2} \left( \frac{\Delta t}{1 ~\mathrm{year}} \right)
\end{equation}
However, as particles are lofted to higher altitudes, they will be surrounded by lower density gas, decreasing the collision rates between the gas molecules and the particle surface.  

Figure 11 shows the cumulative number of collisions that the median micron-sized grain experiences over a 10$^{5}$ year period for the  different $\alpha$ structures defined above.  For comparison, also plotted are the results for those grains which ranked as the 1\% 25\%, 75\% and 99\% grains in terms of collisions experienced over the time interval of interest.  It is important to note that the range of behaviors for the particles is highly dependent on the value or form of $\alpha$.  In all cases, the particles began the simulation at the disk midplane ($z_{0}$=0).

In all cases the average number of collisions experienced by the grains is below that which it would have experienced if it remained at the disk midplane.  This is due to the upward diffusion of the particles into regions of lower density, and thus lower collision rates.  The range of behavior for particles, from those that experienced the minimum and maximum number of collisions is much greater in those cases with low values of $\alpha$.  In such cases, those particles that diffuse upwards early in the simulations can reside in the low density environments for longer periods of time.  Similarly, particles can reside at the disk midplane for long periods of times, being exposed to the maximum collision rate.   This is in contrast to the cases with high-$\alpha$, where the high diffusivity of the disk rapidly cycles all grains through high and low altitudes.  As a result, there is less scatter among the collisional histories of the particles in the high diffusivity simulations, as each grain rapidly approaches the typical value in the simulation. 

Given enough time,  all grains approach a cumulative number of collisions which is $\sim$72\% of the value that they would experience if they resided at the disk midplane throughout in all cases.   This is because the particles and gas are expected to follow gaussian distributions about the disk midplane.  Thus, the average position of the particles is $z$=0.  However, the average distance for particles from the disk midplane, or time averaged height above the disk midplane, would be described by a half-normal distribution (the absolute value of a variable described by a gaussian distribution), which gives an expected value of $\overline{|z|}=\sqrt{\frac{2}{\pi}} \sigma \sim$0.8$H$.  At this height, the gas density is $\sim$72\% the value at the disk midplane, explaining this value that all particles converge to at long times.

Thus, when calculating the average environment to which a dust grain would be exposed in the disk, it is not necessarily the midplane environment that should be used.  In a diffusive environment, particles will be distributed about the midplane with some standard deviation (dust scale-height), $H_{d}$.  The average environment to which the grain would be exposed would thus be at $\overline{|z|} \sim 0.8 H_{d}$.  For large particles that settle efficiently, this location may be very close to the disk midplane (e.g. the centimeter-sized grains in the disk with $\alpha$=10$^{-4}$).  Small grains, or highly diffusive regions may require consideration further away from the disk midplane.  While the drop in gas density is not necessarily large from the disk midplane to $z= \pm 0.8 H$, other effects may combine to create an environment which differs from the disk midplane, such as vertical temperature gradients in the case of viscous heating in the disk, chemical compositional gradients due to differences in the rates of transport with height above the disk midplane, and
the difference in optical depth from the disk surface.

\subsection{Photochemistry}

In addition to gas-solid chemical reactions, solids in primitive materials can be altered through interactions with photons and high-energy particles.  As an example, organic molecules can by synthesized through the UV irradiation of ices \citep{bernstein02,munozcaro02,oberg09}.  Such photons could be part of the interstellar flux, part of the emission from the central star which could be enhanced significantly by the early magnetic activity and flares associated with young T-Tauri stars \citep{feigelson07}, or from nearby massive stars  if the disk exists  within a massive stellar cluster \citep[e.g.][]{adams06,desch07}.  These photons would be incident on the surface of the disk and absorbed by the dust and gas, resulting in  their intensity decaying deeper into the disk.  Thus as dust grains diffuse vertically within a disk they will be exposed to a wide range in the intensity of UV radiation, thus making the level of photochemical processing intimately tied to the paths these particles take within the turbulent disk.

In general, the depth to which UV photons can penetrate in the disk will depend sensitively on the size distribution of dust particles that are present \citep{dd05,ormel07}.  Exceptions are those UV wavelengths which are absorbed directly by gaseous molecules, such as those responsible for the photodissociation of CO or H$_{2}$.    For the purposes of the illustrative examples considered here, we assume that the opacity of the collective gas and dust mixture is constant with a value of $\kappa$=5 cm$^{2}$/g, which corresponds roughly to a collection of dust grains in the gas with an average radius of 15 $\mu$m. A smaller particle size distribution would result in larger opacities, and more rapid attenuation of the UV flux.  As particle growth and settling proceeds, the value of $\kappa$ would decrease with time.

Here micron-sized dust grains in the outer regions of a protoplanetary disk, at $r$=10 AU, are considered.  Values of $\Sigma$=200 g/cm$^{2}$ and $T$=88 K were assumed (taking a surface density that falls off as $r^{-1}$ and a temperature that falls off as $r^{-\frac{1}{2}}$ with the values used at 1 AU in the previous examples).  Temperatures this low would allow water ice to be present as a solid everywhere, and thus such a disk environment is a possible example where organics could be produced by photochemical processes.  Particles were tracked from their release at the disk midplane and the number of photons incident on their surface calculated over time.   The number of photons absorbed in a time interval, $\delta t$, at a position, $z$, is given by:
\begin{equation}
\Delta N_{photons}= I_{0} \pi a^{2} e^{-\kappa \left( \frac{1}{2} \Sigma - \Sigma' \left( z \right) \right)} \delta t
\end{equation}
where $I_{0}$ is the incident flux of photons on the surface of the disk.
An incident flux of $G_{0}$=1 \citep[$I_{0}$=10$^{8}$ UV photons cm$^{-2}$ s$^{-1}$][]{habing68} is assumed. 
The results presented can be scaled to higher UV fluxes  by simply multiplying by the appropriate value of $G_{0}$.  

As in the cases considered above, diffusion works to redistribute grains so that their positions are described by a gaussian about the disk midplane.  Even with the decrease in gas density at 10 AU versus what it was in the cases considered above at 1 AU, the micron-sized grains have a value of $St \sim 6 \times 10^{-7}$ at the disk midplane, meaning they are well coupled to the gas.  Thus, under equilibrium $H_{d} \sim H$ in all cases.  Each of the $\alpha$ profiles discussed previously were investigated.  

Figure 12 shows trajectories of individual grains from each of the four cases considered, along with the cumulative number of photons that were incident on their surfaces.  Again, all particles were released from the disk midplane, where the optical depth is $\tau$=$\frac{1}{2} \kappa \Sigma$=500.  In other words, the incident flux of radiation would be diminished by nearly $\sim$220 orders of magnitude.  Thus particles in the interior of the disk around the midplane would be effectively shielded from the incident radiation, and thus, would not be subjected to photochemical processes.  Under the conditions assumed here, the incident flux will decrease by a factor of 10$^{8}$ (where for $G_{0}$=1, the flux would be equal to 1 photon cm$^{-2}$ s$^{-1}$) at a height of $\Sigma'\left( z \right) \sim$96 g/cm$^{2}$, or $z \sim \pm 2H$.  At higher altitudes, the UV flux would be stronger, and it is at these altitudes where photochemistry is likely to have been important.  

The effect of this photon-rich environment is seen in the bottom panels of particle evolution plots shown in Figure 12.  The photon numbers increase noticeably at the same points where the particles make excursions into the upper layers ($|z|>$2$H$) of the disk.  In the case of the low diffusivity disks ($\alpha$=10$^{-4}$), the number of jumps in the incident photons is fewer, though each jump is relatively large.  This is again due to the slow manner in which particles move about the disk, allowing them to reside in a given environment, and in this case be exposed to a large photon flux, over an extended period of time.  The higher diffusivity disk models (particularly in the upper layers) instead exhibit a larger number of jumps in the incident photons on the particles.  However, these jumps are less dramatic, in that the increase in the number of photons during one excursion to high altitudes is limited as the particles are pulled downward by a strong, effective velocity that pull the grains back toward the disk midplane.  This limits the residence time at high altitudes during any one excursion.

As discussed above, the average particle, given enough time, would spend $\sim$4.5\% of its time in the photon-rich region ($|z|>$2$H$, assuming a gaussian distribution about the disk midplane).  However, such an estimate is valid only after very long timescales.  On shorter timescales, each particle will follow separate paths and be exposed to different dosages of photons.  Figure 13 follows the evolution of the median grain in terms of the number of photons incident on its surface as a function of time for each of the different $\alpha$ cases considered here.  As in the collision results above, the evolution of those particles which ranked 1\%, 25\%, 75\% and 99\% in terms of intercepted photons are also shown.  

Note in Figure 13 that in the cases with $\alpha$=10$^{-4}$  there is no plot for the particles that ranked below the 75\% level in terms of number of photons intercepted.  The reason for this is that in these cases, there are particles that never diffuse to high enough altitudes to be exposed to a significant UV flux.   The higher diffusivity cases have all particles intercepting some photons over the course of the simulation, though there is still orders of magnitude differences between the maximum and minimum exposure at the end of the simulation.  The variations seen here are greater than the variations seen for the number of gas molecule collisions that the grains experience in the previous section.  The reason for this is that while most gas-grain collisions occur around the disk midplane where particles spend most of their time, the UV exposure is determined by the short jumps that the particles experience to high altitudes.  Further, the UV flux decreases very rapidly with depth into the disk, making slight differences in residence times at high altitudes translate into significant differences in UV exposure.  Thus,  the photochemical evolution of primitive materials will be strongly dependent on the paths they take through a protoplanetary disk.

This is particularly true as the chemical evolution that would be caused by the incident radiation is not necessarily solely dependent on the total dose of photons, but rather cycling between the photon-rich and photon-depleted environments.  For example, UV irradiation of C- and N-bearing water ice can lead to the formation of organic molecules, the likes of which have been identified in meteoritic and cometary samples \citep{bernstein02,munozcaro02}.  The basic principle for such reactions is that the UV photons (or other energetic particles) are incident on the particles, breaking chemical bonds and producing ions and radicals.  These photochemical products are able to react with other nearby products, producing more complex molecules.  This synthesis of these more complex molecules can be aided by the warming of the icy substrate \citep{bernstein02,oberg09}, so as to make the materials on the grains more mobile and reactive.  Such warming, would be expected if the particles migrate towards the disk midplane in a viscously heated disk.  Subsequent photoprocessing can lead to more complex organic species.  Thus the efficiency of this mode of production, or the amount of radiation damage and the time that the ions and radicals have to react on the grains, will be determined by the path that each particle takes as it moves around the protoplanetary disk.

\section{Summary}

Here we have developed a model to explore the long-term vertical motions of dust grains in a turbulent protoplanetary disk.  The model is capable of considering the cases where turbulence (diffusivity) is constant with height, as is often assumed, and where turbulence varies with height, as might be expected if the MRI is the primary source of turbulence \citep[e.g.][]{bh91,gammie96,glassgold97,stone01,flemstone03,oishi07}. 
It was shown that while the average long-term behavior of dust particles and gas molecules within a protoplanetary disk is not dependent on the functional form of the diffusivity, the short-term evolution and range in behavior do depend sensitively on how diffusive the disk is and how it varies with location.  In particular, low levels of diffusivity ($\alpha$) result in materials moving slowly from one location to the other, and thus being exposed to a given environment continuously for long periods of time, particularly around the disk midplane.  Highly diffusive disks, on the other hand, quickly cycle materials through different locations, exposing them to a large number of different environments, but each one only very briefly at any given time.  

The different types of paths that materials would take in a protoplanetary disk would lead to different levels and types of chemical and physical processing.  In terms of chemical evolution, the types and rates of chemical reactions would vary with the different environments that dust grains would be exposed to over the course of their lifetimes in the disk.  That is, chemical reactions between primitive solids and gas molecules will be determined in part by the pressure of the gas surrounding the solids, which varies with height in a protoplanetary disk.  Further, the flux of energetic photons or particles will vary with height throughout the disk, being more intense near the disk surface and dropping rapidly towards the disk midplane.  Thus particular reactions would be limited to occur within particular vertical environments within a disk, and it is the motions of the particles through the disk which will determine which reactions take place and to what extent.
In the case of low-diffusivity, as mentioned above, materials will move slowly from one environment to the next.  As a result, the chemical evolution of materials within the disk would exhibit a much wider range of outcomes than in a disk with a high level of diffusivity.  This was shown in Figures 11 and 13 where the ranges of collisions experienced by dust grains and ranges of photons absorbed by those grains within different diffusive environments were shown for the different cases considered.

The methods outlined here provide a means of quantifying the types of trajectories that particles would take within diffusive protoplanetary disks.  This method of calculating the motions of the grains can be used to extend shearing box calculations which track dust grain motions in MRI-driven turbulence, as the length of time that can be simulated is limited to $< 100$ orbits in such models \citep[e.g.][]{turner10}.  As the lifetimes of grains and the different alteration reactions of primitive materials in protoplanetary disks can have timescales that exceed this period of time, the method outlined here allows longer time periods and more particles to be followed, provided the spatial dependence of diffusivity is known.  Further, with the development of Monte Carlo coagulation codes \citep{ormel08,zsom08} the methods developed here can be adopted to provide a way of tracking the chemical evolution of primitive materials and their incorporation into the building blocks of planets.   Indeed, it is this long-term behavior that will be critical in fully understanding how the primitive materials we observe in meteorites, comets, and in disks around other stars developed their properties.

\emph{Acknowledgments} The author is grateful for comments and suggestions offered by Chris Ormel in his careful review of the original version of this manuscript. Conversations with Jeff Cuzzi and Scott Sandford were also helpful in the development of this work.  This work was supported by NASA Grant NNX08AY47G awarded to FJC.

\section{Appendix}

Monte Carlo techniques such as those used here have been developed and applied to address a number of issues in the Earth Sciences.  As such, a variety of approaches have been developed in order to formulate the proper technique for tracking particles in turbulent environments.  Here an alternative approach to deriving the method of treating particle motions in environments with spatially varying turbulence is presented following the methods of \citet{hunter93} and \citet{visser97}.

The diffusion equation with a spatially varying diffusivity is written as:
\begin{equation}
\frac{\partial f}{\partial t} = \frac{\partial}{\partial z} \left( D_{z} \frac{\partial f}{\partial z} \right) 
\end{equation}
As $D_{z}$ is a function of location, we can write it of the form:
\begin{equation}
D_{z} = D_{0} + z D'_{0}
\end{equation}
where $D_{0}$=$D_{z} \left( 0 \right)$, $D'_{0}$=$\left( \frac{\partial D_{z}}{\partial z }\right)_{z=0}$, and $z=0$ represents the current location of the particle.  Substituting this into the diffusion equation gives:
\begin{equation}
\frac{\partial f}{\partial t} = D_{0} \frac{\partial^{2} f}{\partial z^{2}} + z D'_{0} \frac{\partial^{2} f}{\partial z^{2}} + D'_{0} \frac{\partial f}{\partial z}
\end{equation}

The moments of a distribution can provide information about how the entire ensemble of particles evolves with time.
We can define the $n-th$ moment of the distribution of the materials as:
\begin{equation}
M_{n} = \int_{-\infty}^{\infty} z^{n} f dz
\end{equation}
With this definition, the zeroth moment ($n$=0) would provide the total mass of the distribution, the first moment ($n$=1) would provide the center of mass of the distribution, the second moment ($n$=2) provides information on the variance of the distribution.

Differentiating allows us to evaluate how the distribution evolves with time:
\begin{equation}
\frac{\partial M_{n}}{\partial t} =  \int_{-\infty}^{\infty} z^{n} \frac{\partial f}{\partial t} dz
\end{equation}
Substituting into the above equation gives:

\begin{equation}
\frac{d M_{n}}{d t} = 
\int_{-\infty}^{\infty} z^{n} \frac{\partial f}{\partial t} dz= 
\int_{-\infty}^{\infty} z^{n} D_{0} \frac{\partial^{2} f}{\partial z^{2}} dz + 
\int_{-\infty}^{\infty} z^{n+1} D'_{0} \frac{\partial^{2} f}{\partial z^{2}} dz + 
\int_{-\infty}^{\infty} z^{n} D'_{0} \frac{\partial f}{\partial z} dz
\end{equation}

In order to understand how the moments evolve, it is necessary to evaluate the above integrals to come up with a functional form for $M_{n}$.  This can be done by recognizing both $f \rightarrow$0 and $\frac{\partial f}{\partial z} \rightarrow$0 as $z \rightarrow \pm \infty$.  Taking the first term on the right hand side of the equation above, integration by parts yields:
\begin{equation}
\int_{-\infty}^{\infty} z^{n} D_{0} \frac{\partial^{2} f}{\partial z^{2}} dz = 
\left[D_{0} z^{n} \frac{\partial f}{\partial z} \right]_{-\infty}^{\infty} - D_{0} \int_{-\infty}^{\infty} n z^{n-1} \frac{\partial f}{\partial z} dz
\end{equation}
Because of the limits of $\frac{\partial f}{\partial z}$ identified above, the first term on the right is equal to zero.  Integration by parts of the second term yields:
\begin{equation}
- D_{0} \int_{-\infty}^{\infty} n z^{n-1} \frac{\partial f}{\partial z} =
-\left[ D_{0} n z^{n-1} f \right]_{-\infty}^{\infty} + D_{0} n \left( n-1 \right) \int_{-\infty}^{\infty} z^{n-2} f dz
\end{equation}
Again the first term on the right is equal to zero due to the limits defined above.  This means that the first term on the right of Equation (38) reduces to:
\begin{equation}
\int_{-\infty}^{\infty} z^{n} D_{0} \frac{\partial^{2} f}{\partial z^{2}} dz =
D_{0} n \left( n-1 \right) \int_{-\infty}^{\infty} z^{n-2} f dz =
D_{0} n \left( n-1 \right) M_{n-2}
\end{equation}

A similar approach for the other terms in Equation (38) gives:
\begin{equation}
\int_{-\infty}^{\infty} z^{n+1} D'_{0} \frac{\partial^{2} f}{\partial z^{2}} dz =
D'_{0} n \left( n+1 \right) M_{n-1}
\end{equation}
and
\begin{equation}
\int_{-\infty}^{\infty} z^{n} D'_{0} \frac{\partial f}{\partial z} dz =
-D'_{0} n M_{n-1}
\end{equation}

Putting these terms together, we finally get:
\begin{equation}
\frac{d M_{n}}{d t} = D_{0} n \left( n - 1 \right) M_{n-2} + D'_{0} n^{2} M_{n-1}
\end{equation}

As the zeroth moment, $M_{0}$=$\int_{-\infty}^{\infty} f dz$, is simply the total mass (surface
density in the cases considered in this work) of the species , which is constant ($\frac{d M_{n}}{d t}$=0), we can define a series of normalized moments, $N_{n}$ such that:
\begin{equation}
N_{n} = \frac{M_{n}}{M_{0}}
\end{equation}
Combining Equations (44) and (45) allows us to write the time derivatives of the first and second normalized moments  as:
\begin{equation}
\frac{d N_{1}}{d t} = D'_{0}
\end{equation}
\begin{equation}
\frac{d N_{2}}{d t} = 2 D_{0} + 4 D'_{0}N_{1}
\end{equation}

Again, the first normalized moment represents the center of mass of a distribution.  In simple cases, when the diffusivity of materials is constant and in the absence of advection, materials starting at some $z=z_{0}$ will diffuse away from the starting location such that the variance grows as $\sigma^{2}$=$2Dt$, however the average position, or center of mass, would remain fixed at $z=z_{0}$.  With a uniform diffusivity here, $D'_{0}=0$, and the predicted evolution here is that the center of mass remains fixed with time, or $d N_{1}$/$dt$=0.  If spatial gradients in the diffusivity do exist, however, $D'_{0}$ would have some non-zero value, meaning that diffusion would result in the migration of the center of mass of the distribution in the direction of positive $D'_{0}$, or increasing diffusivity.

The second normalized moment describes the variance of the distribution, and as can be seen if $D'_{0}=0$, we get the expected result that the rate of change of the variance ($\sigma^{2}$) with time is given by $2D_{0}$.  However, with a spatially varying diffusivity, Equation (47) shows that the variance will have a more complicated evolution.  To illustrate this, consider the case of a collection of materials beginning at $z=0$ at $t=0$ and diffusing in a medium in which the diffusivity is not constant.  After a time, $\delta t$, the change in the first normalized moment is:
\begin{equation}
\int_{0}^{\delta t} \frac{d N_{1}}{d t} dt = N_{1} \left( \delta t \right) - N_{1} \left( 0 \right) = D'_{0} \delta t
\end{equation}

In other words, after a time period, $\delta t$, the center of mass would have migrated to $z$=$D'_{0} \delta t$.  For the second moment:
\begin{equation}
\int_{0}^{\delta t} \frac{d N_{2}}{d t} dt = \int_{0}^{\delta t} 2 D_{0} dt + \int_{0}^{\delta t} 4 D'_{0} N_{1} dt
\end{equation}
The first term on the right simply gives:
\begin{equation}
 \int_{0}^{\delta t} 2 D_{0} dt  = 2 D_{0} \delta t
 \end{equation}
 which is again the expected result in the case of $D'_{0}$=0.  
 
 The second term on the right of Equation (49), however requires some manipulation:
 \begin{equation}
 \int_{0}^{\delta t} 4 D'_{0} N_{1} dt = \left[4 D'_{0} N_{1} \left( \delta t \right) \right]_{0}^{\delta t} - 4 D'_{0} \int_{0}^{\delta t} t \frac{d N_{1}}{d t} dt
 \end{equation}
 using the fact that $\frac{d N_{1}}{d t}$=$D'_{0}$ allows us to simplify:
 \begin{equation}
 \int_{0}^{\delta t} 4 D'_{0} N_{1} dt = 4 D'_{0} N_{1} \left( \delta t \right) - 4 D'_{0}  \int_{0}^{\delta t} t D'_{0} dt
 \end{equation}
 The second term on the right becomes:
 \begin{equation}
4 D'_{0}  \int_{0}^{\delta t} t D'_{0} dt = \left[ 2 D_{0}^{'2} t^{2} \right]_{0}^{\delta t}
 \end{equation}
 or
 \begin{equation}
4 D'_{0}  \int_{0}^{\delta t} t D'_{0} dt = 2 D_{0}^{'2}  \delta t^{2} 
 \end{equation}
Recall from above:
\begin{equation}
D'_{0} \delta t = N_{1} \left( \delta t \right)
\end{equation}
which means that the second term on the right of Equation (52) becomes:
\begin{equation}
4 D'_{0}  \int_{0}^{\delta t} t D'_{0} dt = 2 D'_{0} N_{1} \left( \delta t \right) \delta t
\end{equation}
which we do not simplify further so that this expression is easily used further below.  Plugging this into Equation (52) gives:
\begin{equation}
\int_{0}^{\delta t} 4 D'_{0} N_{1} dt = 4 D'_{0} N_{1} \left( \delta t \right) \delta t - 2 D'_{0} N_{1} \left( \delta t \right) \delta t 
 = 2 D'_{0} N_{1} \left( \delta t \right) \delta t 
\end{equation}  
Thus Equation (49) gives:
\begin{equation}
\int_{0}^{\delta t} \frac{d N_{2}}{d t} dt = N_{2} \left( \delta t \right) - N_{2} \left( 0 \right) = 2 D_{0} \delta t + 2 D'_{0} N_{1} \left( \delta t \right) \delta t 
\end{equation}
which simplifies to:
\begin{equation}
N_{2} \left( \delta t \right) = 2 \left( D_{0} + D'_{0} N_{1} \left(\delta t \right) \right) \delta t 
\end{equation}

Now the expression found for $N_{2} \left( \delta t \right)$ is by definition, the mean squared distance from the origin of the coordinate system.  Here it is shown that $N_{2}$ would thus be equivalent to 2 times the timestep, $\delta t$, times the value of the diffusivity evaluated at $z$=$N_{1} \left(\delta t \right)$.
However, the variance of the distribution of particles would be the mean squared distance from the center of mass of the particles, which after a timestep $\delta t$ is located at $z=N_{1} \left( \delta t \right)$.  This gives:
\begin{equation}
N_{2} \left( \delta t \right) - N_{1}^{2} \left( \delta t \right) = 2 \left( D_{0} \delta t + D'_{0} N_{1} \left( \delta t \right) \right) \delta t - D'_{0} N_{1} \left( \delta t \right) \delta t 
\end{equation}
or
\begin{equation}
N_{2} \left( \delta t \right) - N_{1}^{2} \left( \delta t \right) = 2 \left( D_{0} + \frac{1}{2} D'_{0}  N_{1} \left( \delta t \right) \right) \delta t 
\end{equation}
Note that $D_{0} + \frac{1}{2} N_{1} \left(\delta t \right) D'_{0} \delta t$ is simply the value of $D_{z}$ evaluated at $z$=$ \frac{1}{2} N_{1} \left( \delta t \right)$, and thus
\begin{equation}
N_{2} \left( \delta t \right) - N_{1}^{2} \left( \delta t \right) = 2 D \left[ \frac{1}{2} N_{1} \left( \delta t \right) \right] \delta t
\end{equation}

In developing a Monte Carlo model for diffusion with a spatially varying diffusion coefficient, we must develop a model which allows the collection of particles to evolve in a manner such that the moments of the center of mass and variance of the distribution are consistent with the above derivations.  In order to do so, an advective velocity is added which equals the spatial gradient in the diffusivity, so that all particles will migrate in the direction of increasing diffusivity, consistent with Equation (48).

The random displacements due to diffusion must, on average, increase with time as particles drift towards regions of higher diffusivity.  That is, the random displacement in the case of constant diffusivity would be described by: 
\begin{equation}
\Delta z_{rand}^{const} = R \left( \frac{2}{r} D \delta t \right)^{\frac{1}{2}}
\end{equation}
As discussed above, the variance should be greater than 2$D_{0} \delta t$ after a timestep, $\delta t$, in the presence of spatially varying diffusivity, which would not be the case if we simply evaluated the random displacement at the starting point of the particles.  Instead, the random displacement should be evaluated at a position offset by some distance in the direction of increasing diffusivity so that its value is consistent with Equation (63).  As such, \citet{visser97} suggested using:
\begin{equation}
\Delta z_{rand}^{var} = R \left[ \frac{2}{r} D \left( z' \right) \delta t \right]^{\frac{1}{2}}
\end{equation}
where 
\begin{equation}
z'=z_{i-1} + \frac{1}{2} D'_{0} \left( z_{i-1} \right) \delta t
\end{equation}
and $z_{i-1}$ is the location of the particle at the beginning of the timestep.

Putting this all together, the position, $z_{i}$ of a particle in a Monte Carlo simulation of transport in a spatially varying diffusivity, beginning at position $z_{i-1}$, after a timestep $\delta t$ would be given by:
\begin{equation}
z_{i} = z_{i-1} +  D'_{0} \left( z_{i-1} \right) \delta t + R \left[ \frac{2}{r} D \left( z' \right) \delta t \right]^{\frac{1}{2}}
\end{equation}
onto which further advective motions can be added.

\bibliographystyle{apj}
\bibliography{rwalk}

\newpage
\begin{figure}
\includegraphics[angle=90,width=5in]{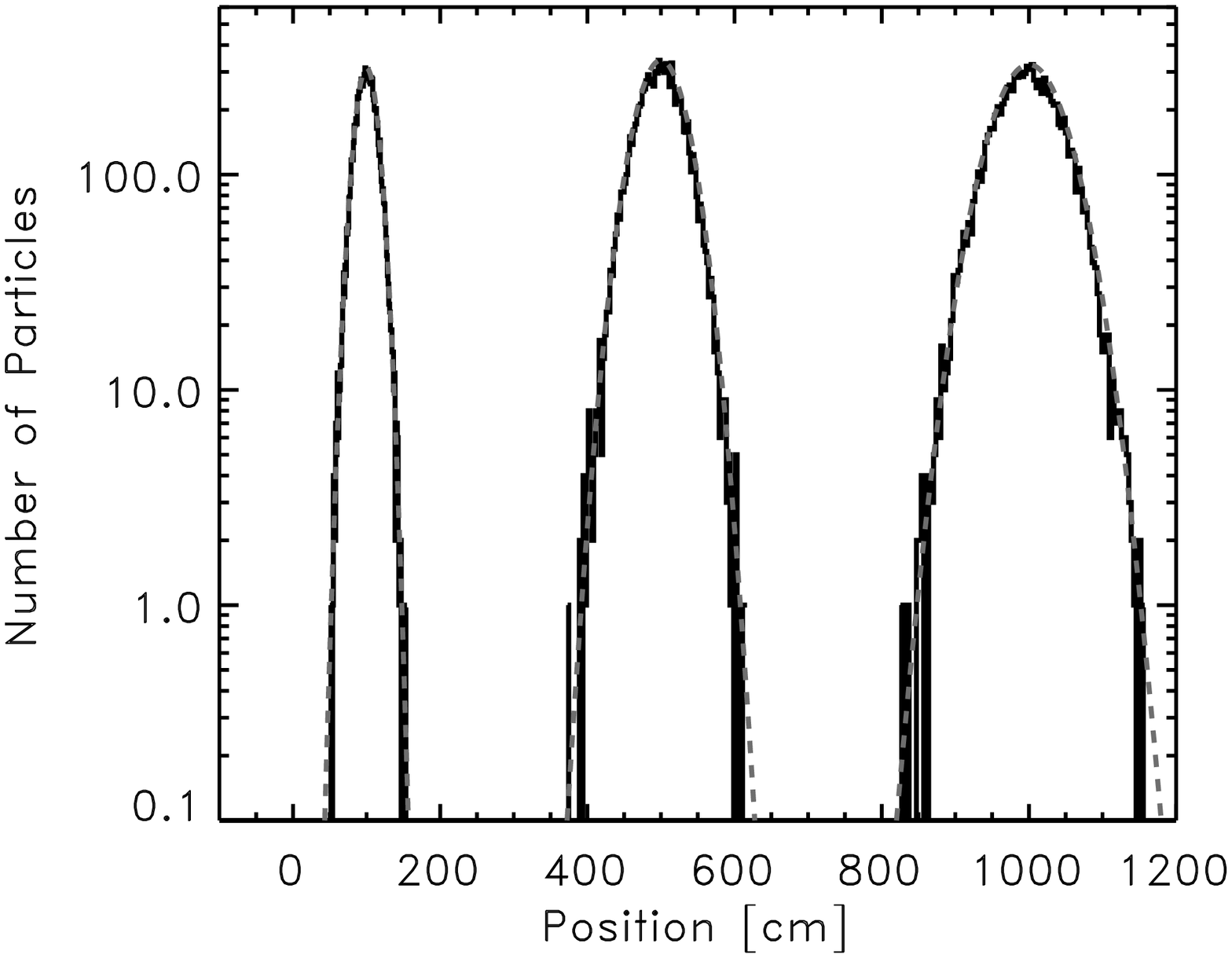}
\caption{Histograms (black lines) of the positions of 10,000 particles subjected to an advective velocity of 10 cm/s and a diffusivity of 10 cm$^{2}$/s at $t$=10, 50, and 100 seconds.   Grey curves show the analytic solutions for comparison. These agreements serve to validate the model used here.}
\end{figure}

\newpage
\begin{figure}
\includegraphics[angle=90,width=5in]{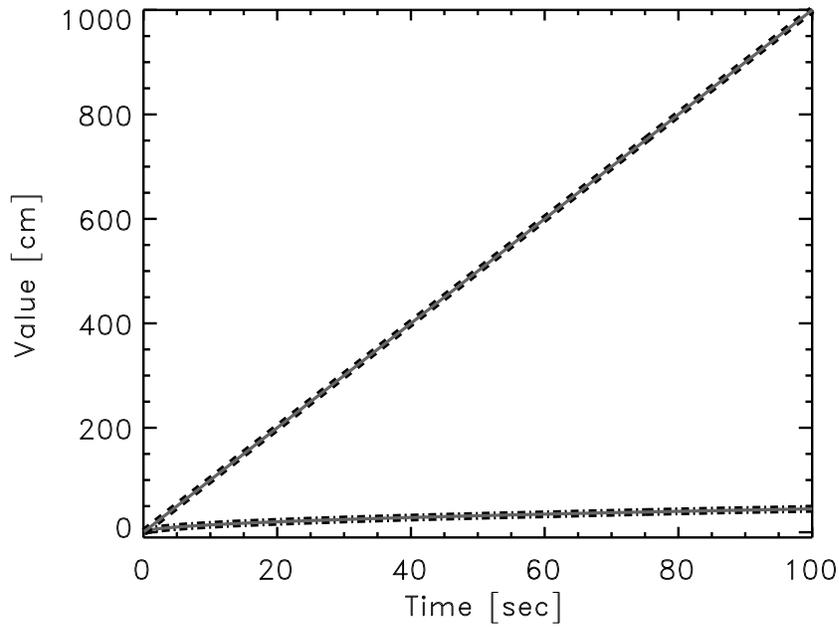}
\caption{Comparison of the temporal evolution of the mean position (dashed line) and standard deviation (dash-dot line) of the particles in the Monte Carlo simulation of the advection-diffusion equation with analytic solutions (grey lines). These agreements serve to validate the model used here.}
\end{figure}

\newpage
\begin{figure}
\includegraphics[angle=90,width=5in]{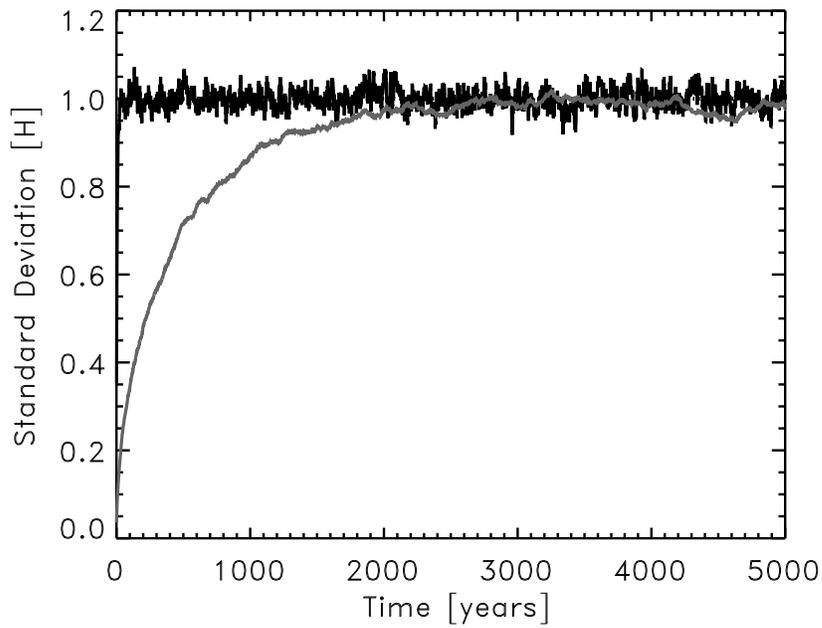}
\caption{Temporal evolution of the standard deviation of the positions of gaseous molecules released at the midplane of a disk at 1 AU with a local surface density of $\Sigma$=2000 g/cm$^{2}$ and $T$=280 K.  The black line represents the evolution of molecules in a disk with $\alpha$=10$^{-2}$ while grey is for the case of $\alpha$=10$^{-4}$.}
\end{figure}

\newpage
\begin{figure}
\includegraphics[angle=90,width=5in]{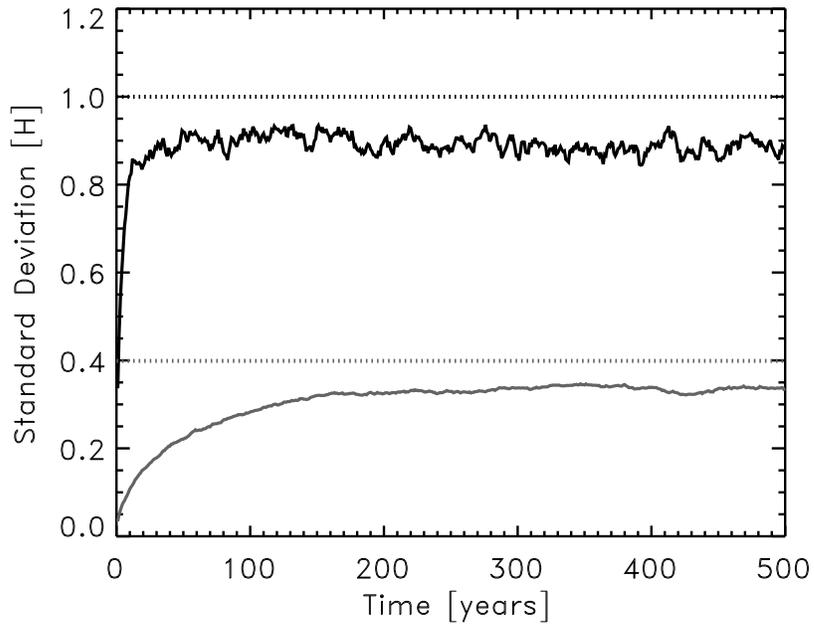}
\caption{Temporal evolution of the standard deviation of the positions of centimeter-sized particles released at the midplane of a disk at 1 AU for the same conditions in Figure 3.  The black line represents the evolution of molecules in a disk with $\alpha$=10$^{-2}$ while grey is for the case of $\alpha$=10$^{-4}$. The two dashed lines at $z=H$ and $z=0.4H$ represent the analytic predictions for the standard deviations for these two cases respectively.}
\end{figure}

\newpage
\begin{figure}
\includegraphics[angle=90,width=5in]{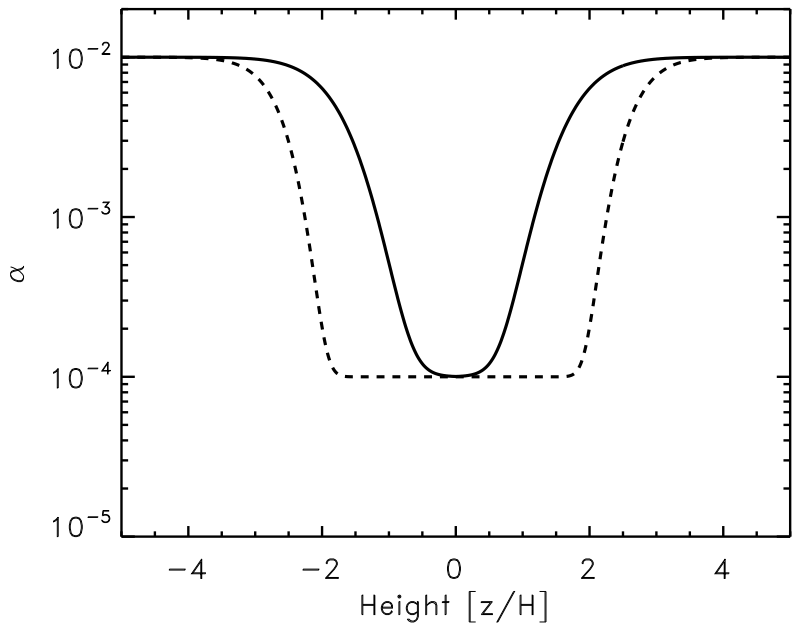}
\caption{The variations of $\alpha$ with height using the expression in Equation (25) for the case of $\Sigma_{c}$=100 g/cm$^{2}$ (solid line) and $\Sigma_{c}$=10 g/cm$^{2}$ (dashed line) at 1 AU for the conditions described in Figure 3. }
\end{figure}

\newpage
\begin{figure}
\includegraphics[angle=90,width=5in]{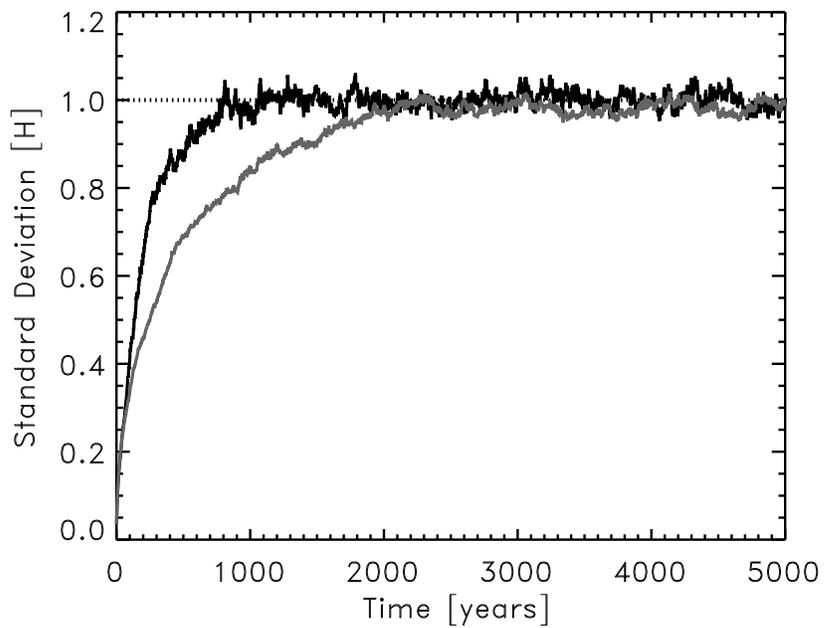}
\caption{Temporal evolution of the standard deviation of the positions of gaseous molecules released at the midplane of a disk for a disk with the $\alpha$ structures shown in Figure 5.  The black line is for the case of $\Sigma_{c}$=100 g/cm$^{2}$ while the grey line is for $\Sigma_{c}$=10 g/cm$^{2}$.}
\end{figure}

\newpage
\begin{figure}
\includegraphics[angle=90,width=5in]{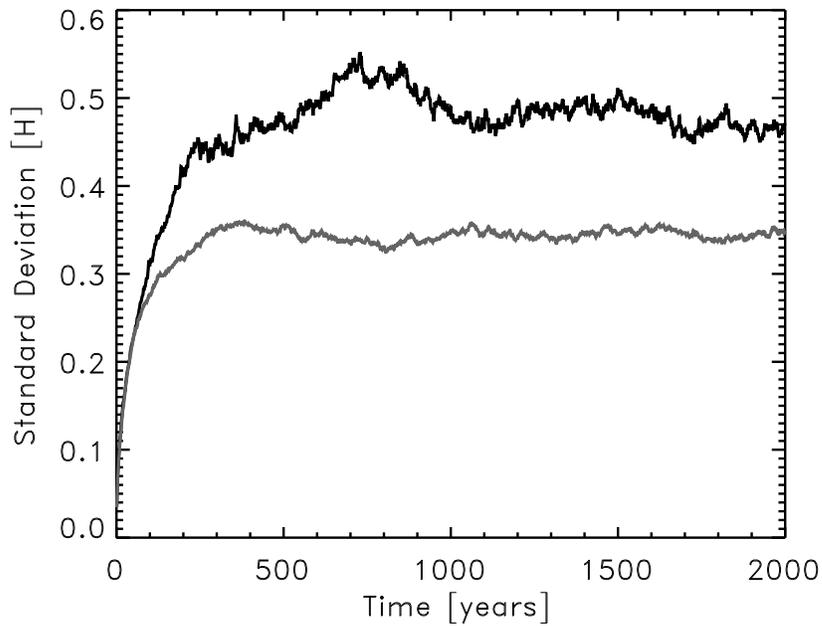}
\caption{Temporal evolution of the standard deviation of the positions of centimeter-sized particles released at the midplane of a disk for a disk with the $\alpha$ structures shown in Figure 5.  The black line is for the case of $\Sigma_{c}$=100 g/cm$^{2}$ while the grey line is for $\Sigma_{c}$=10 g/cm$^{2}$.}
\end{figure}

\newpage
\begin{figure}
\includegraphics[angle=90,width=3in]{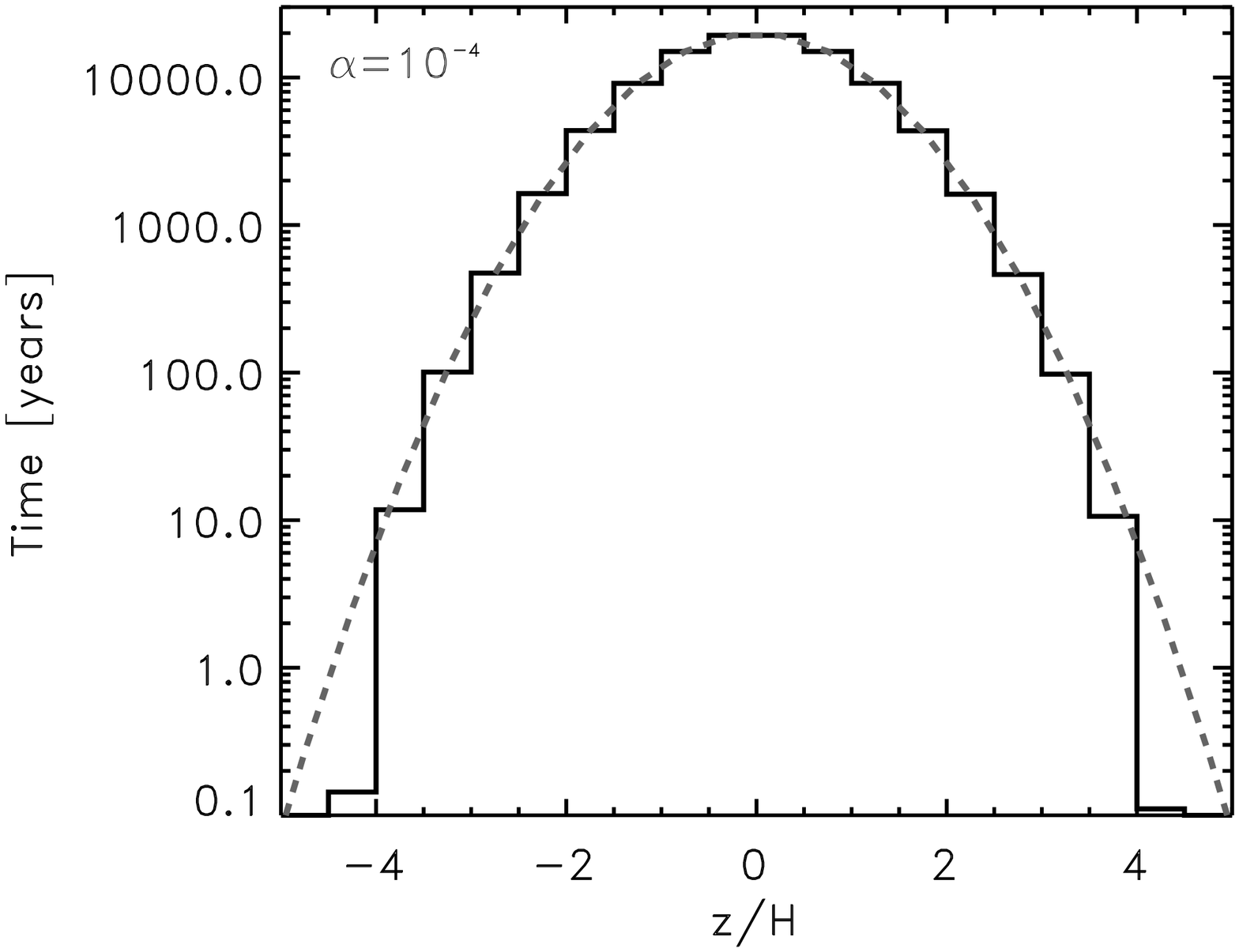}
\includegraphics[angle=90,width=3in]{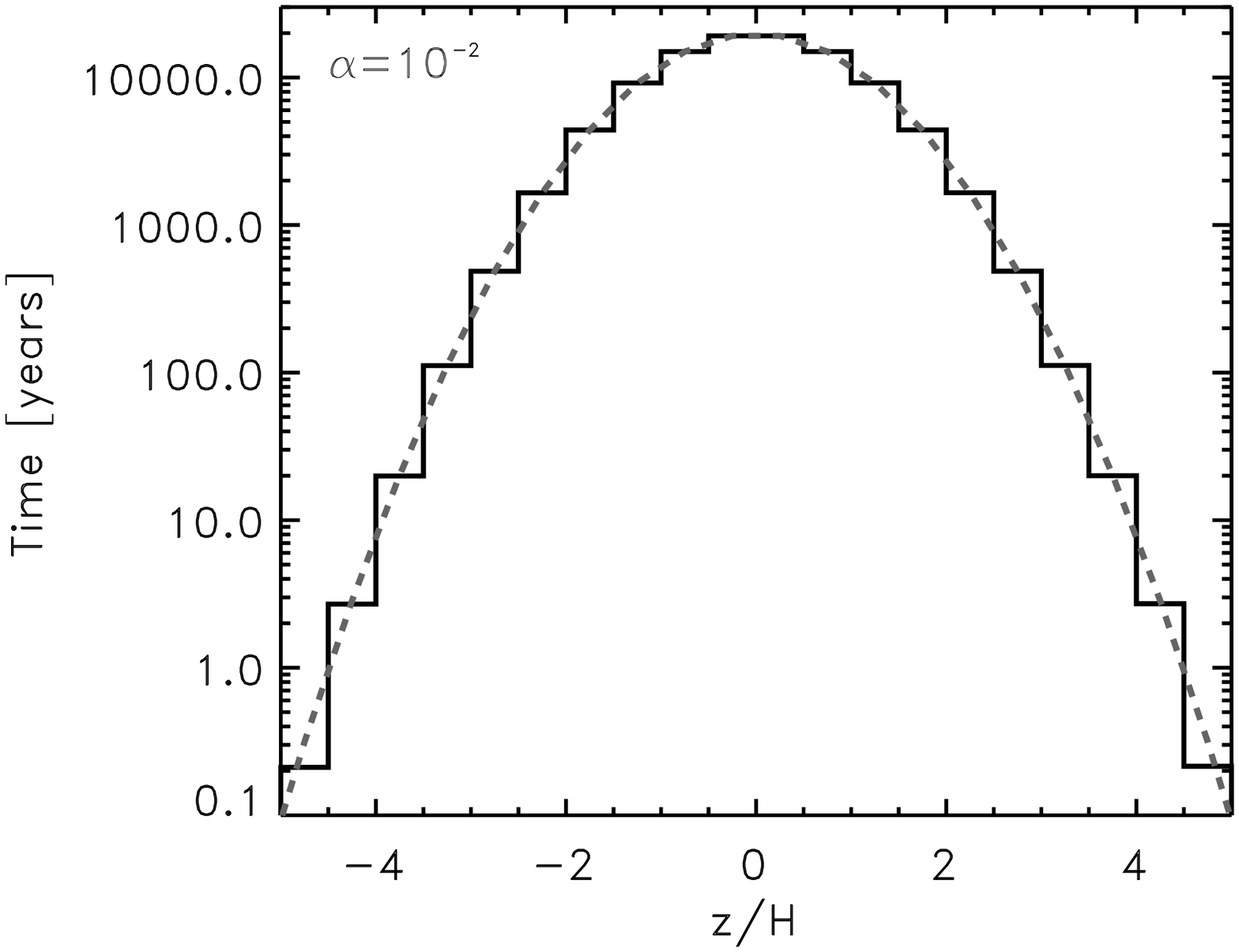}
\includegraphics[angle=90,width=3in]{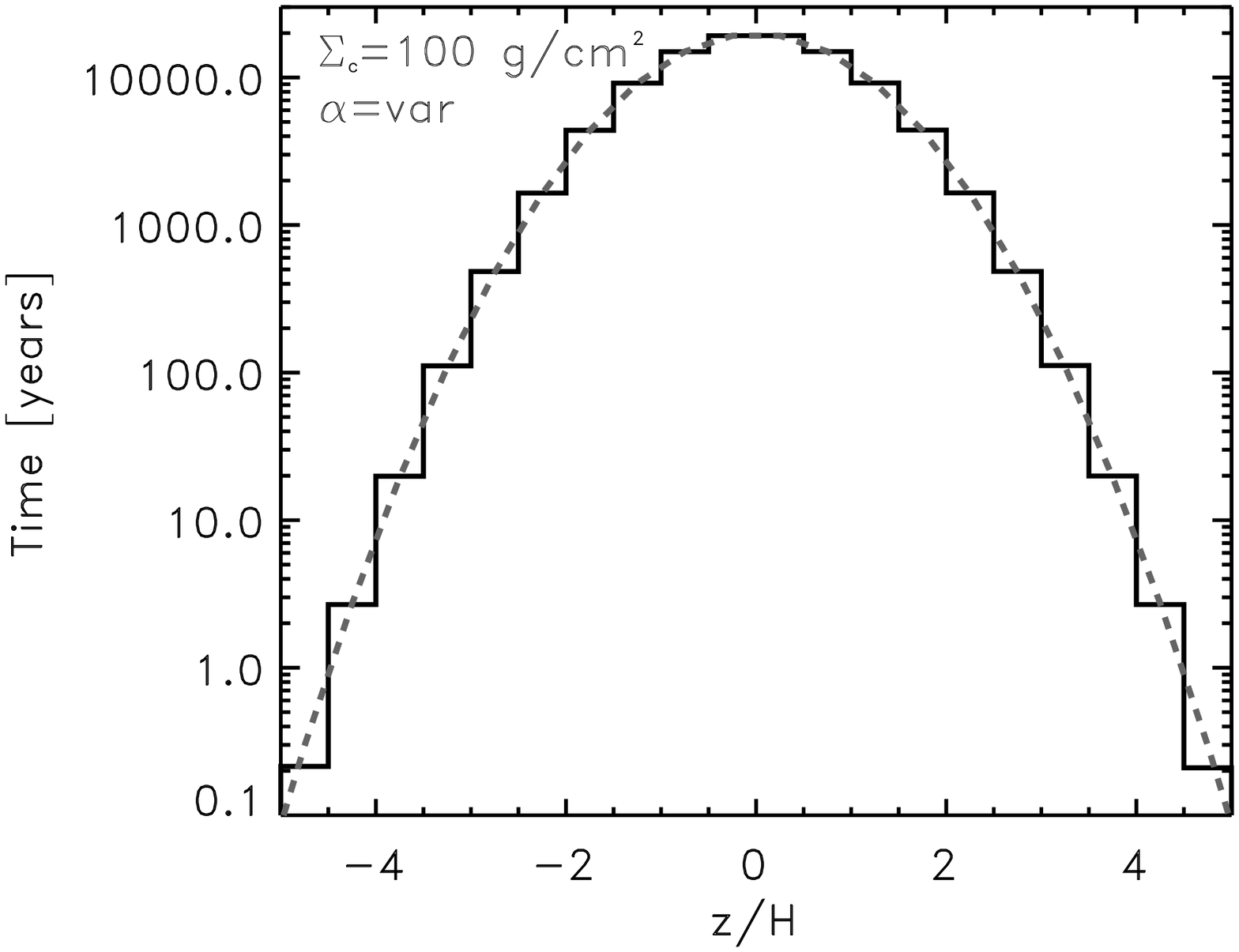}
\includegraphics[angle=90,width=3in]{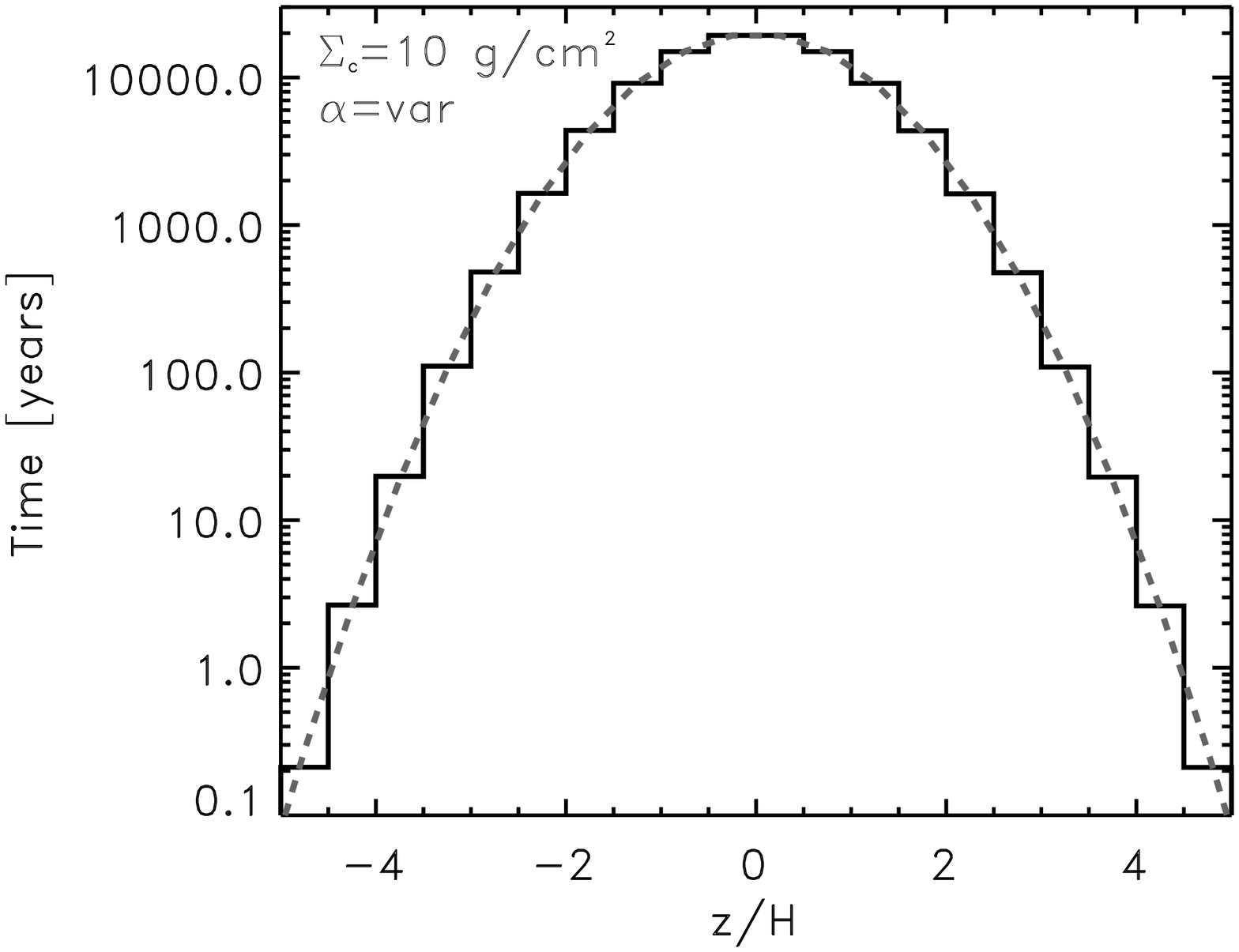}
\caption{Histograms for the the time spent by micron-sized particles  over a 10$^{5}$ year time period at different heights in the disk for the various $\alpha$ cases considered in this study.  In all cases, the histograms match well with a gaussian function with a standard deviation of $\sigma$=$H$ (dashed-grey line in each plot).}
\end{figure}

\newpage
\begin{figure}
\includegraphics[angle=90,width=3in]{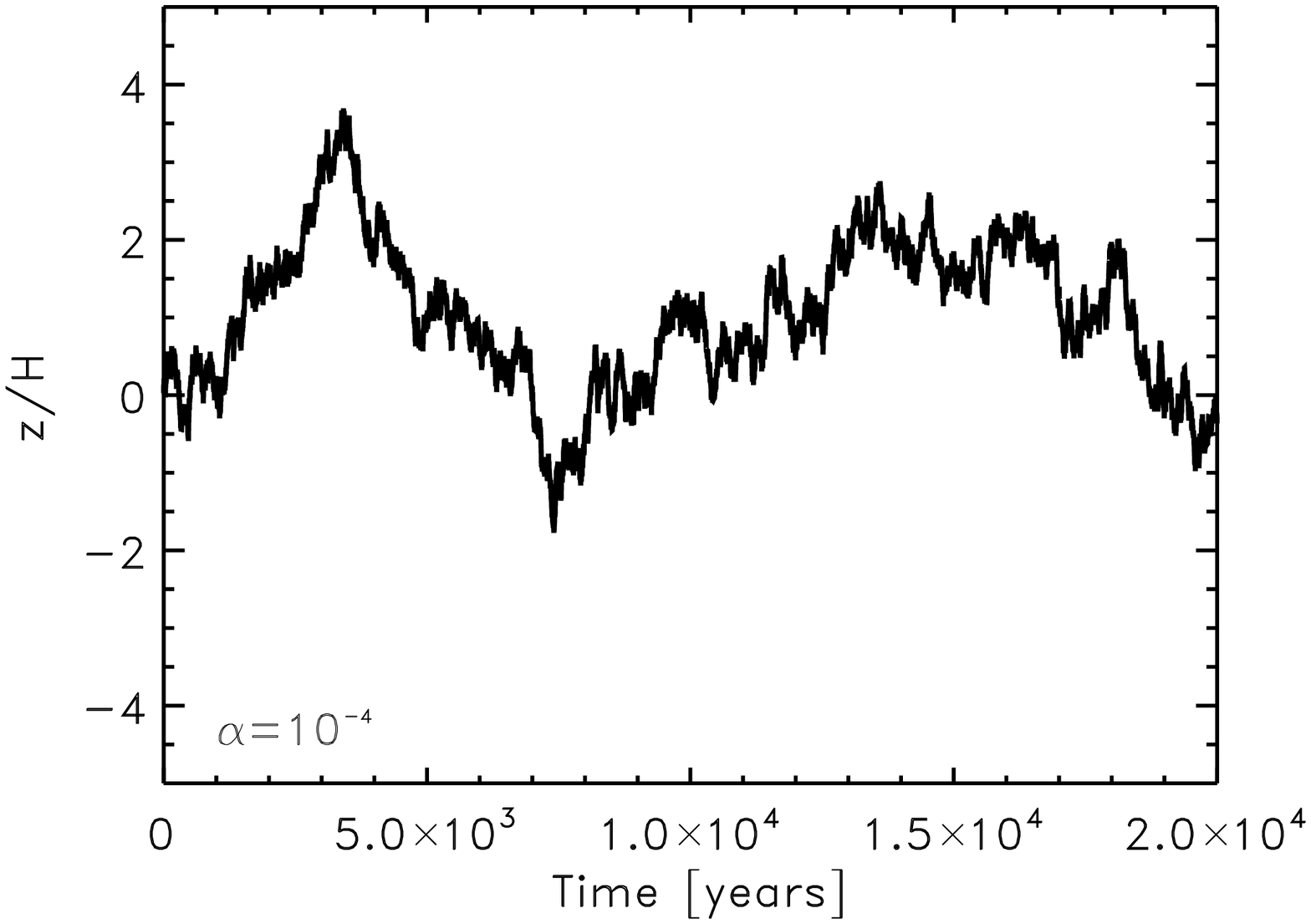}
\includegraphics[angle=90,width=3in]{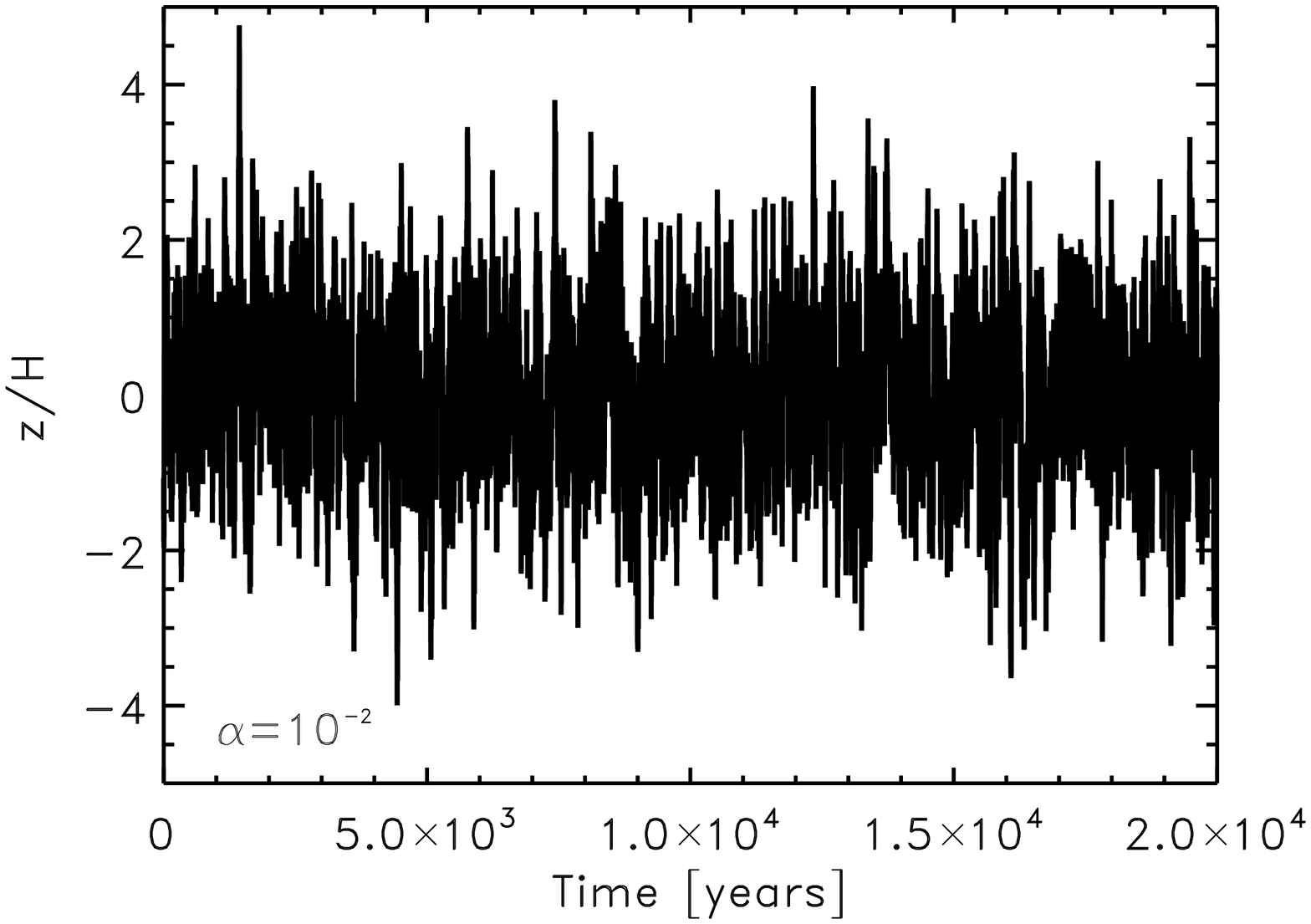}
\includegraphics[angle=90,width=3in]{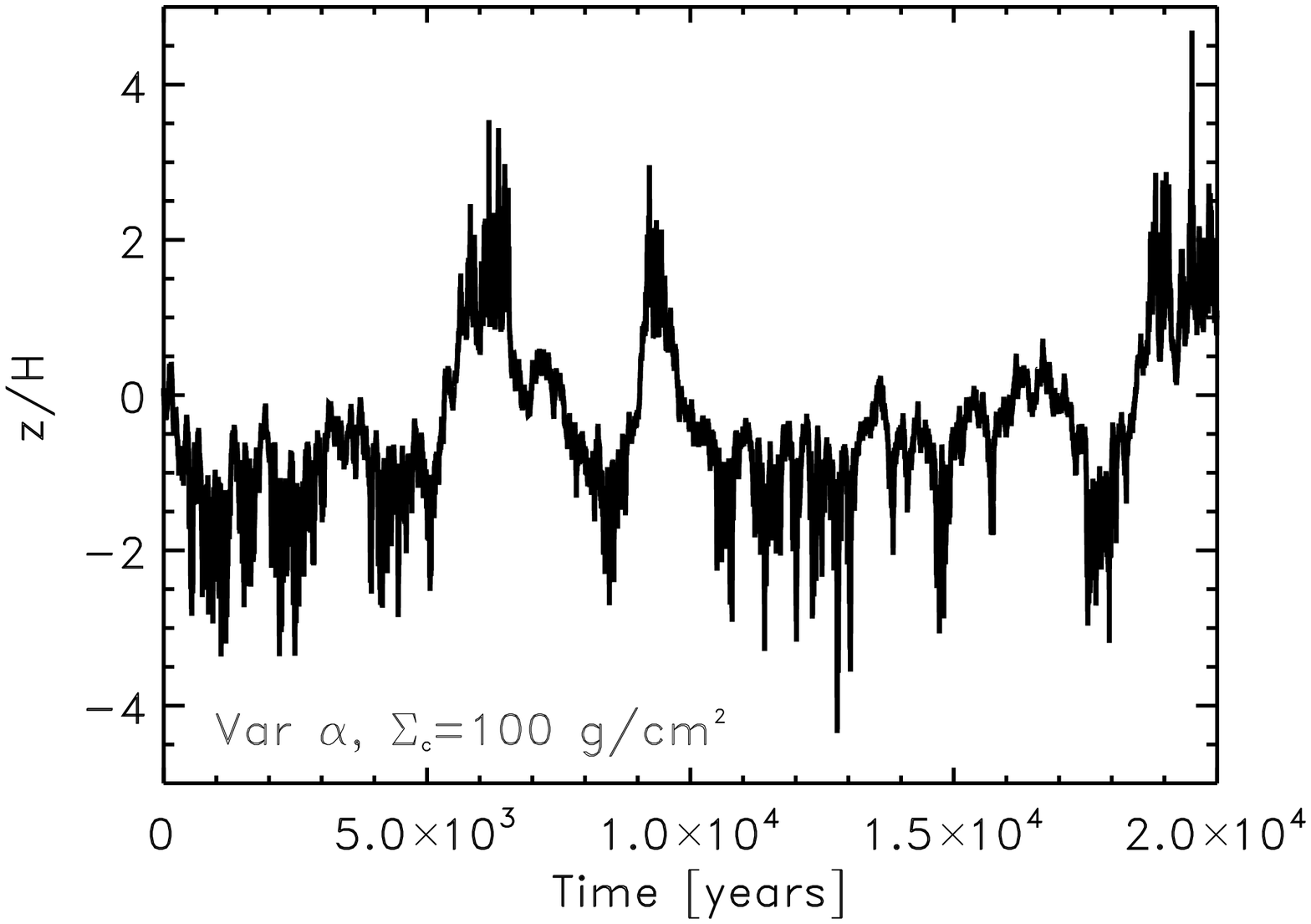}
\includegraphics[angle=90,width=3in]{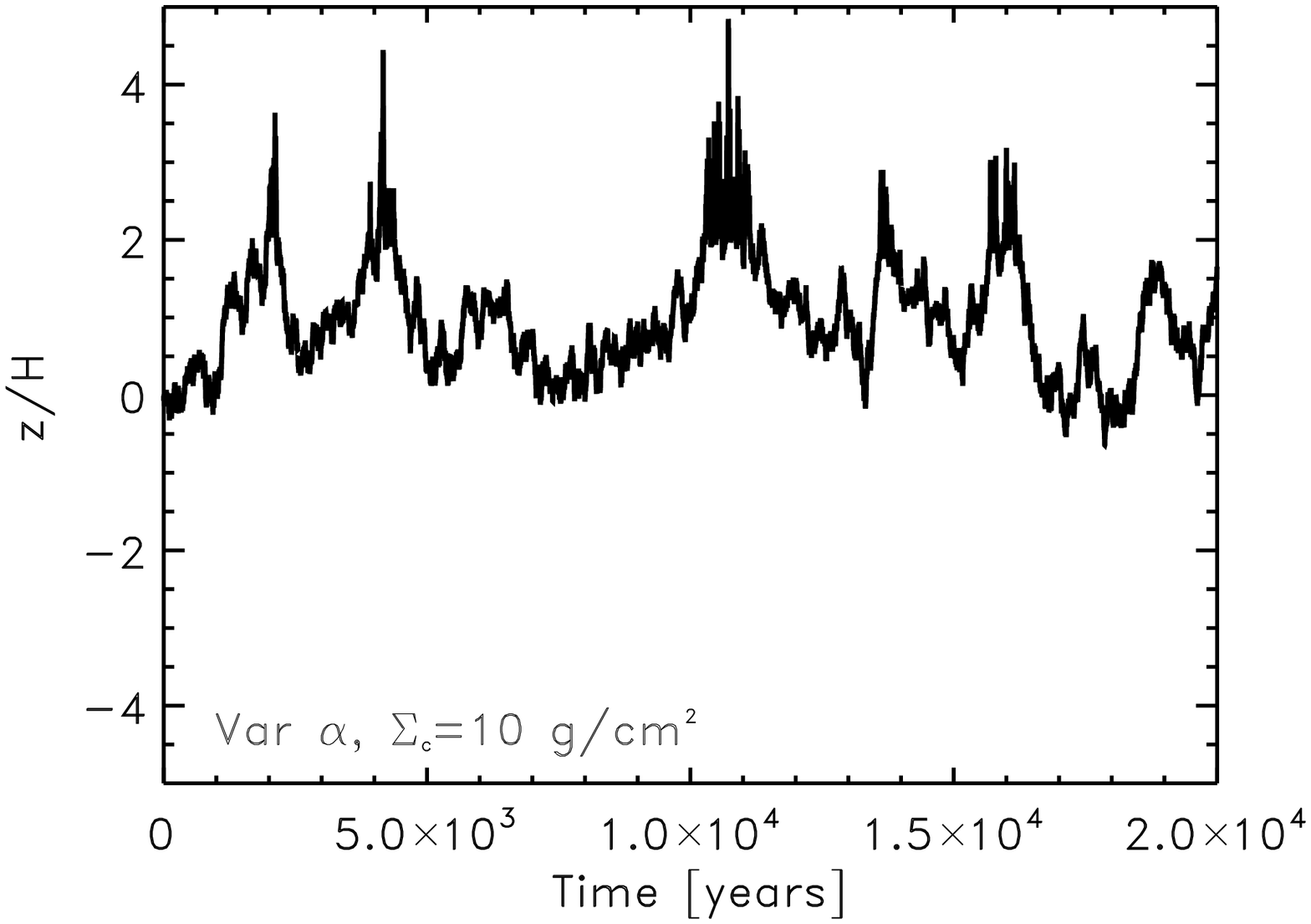}
\caption{Trajectories for individual grains over 20,000 year time periods for the different $\alpha$ cases considered in this study.  In regions of low $\alpha$, particles slowly bounce around, spending long periods of time in a given environment.  In high $\alpha$ regions, particles experience large excursions on short time periods, leading to the introduction to new environments very rapidly.}
\end{figure}

\newpage
\begin{figure}
\includegraphics[angle=90,width=3in]{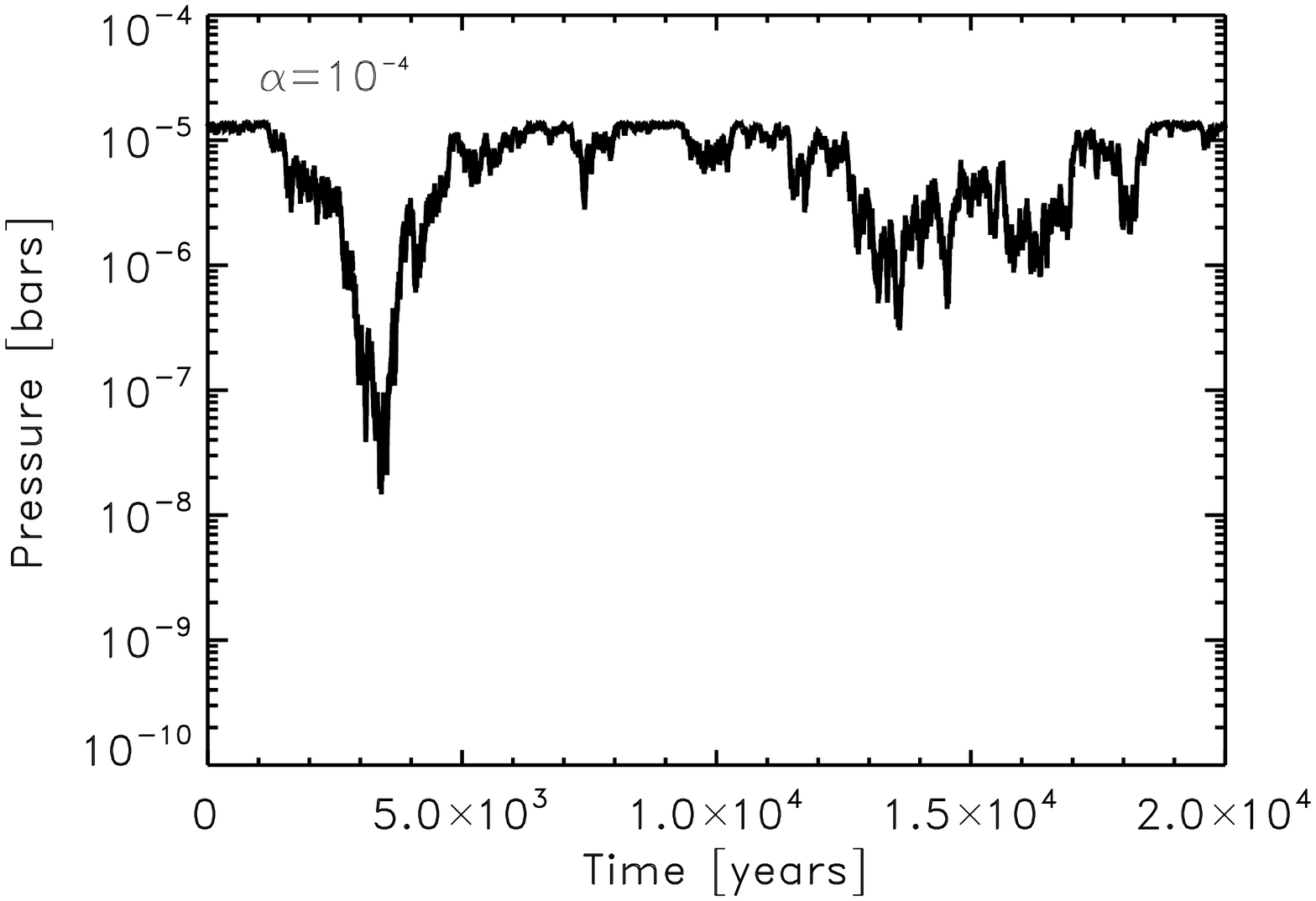}
\includegraphics[angle=90,width=3in]{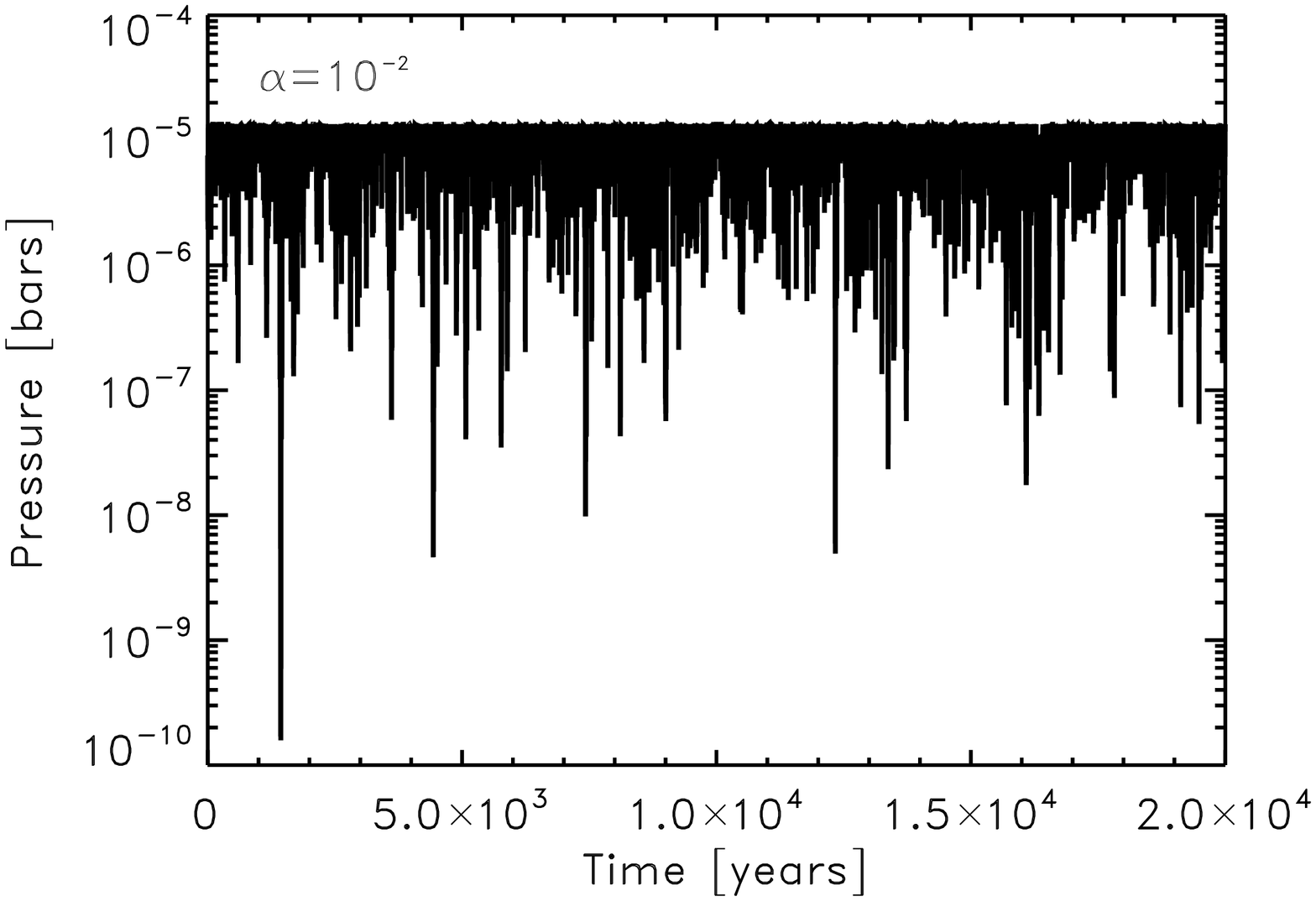}
\includegraphics[angle=90,width=3in]{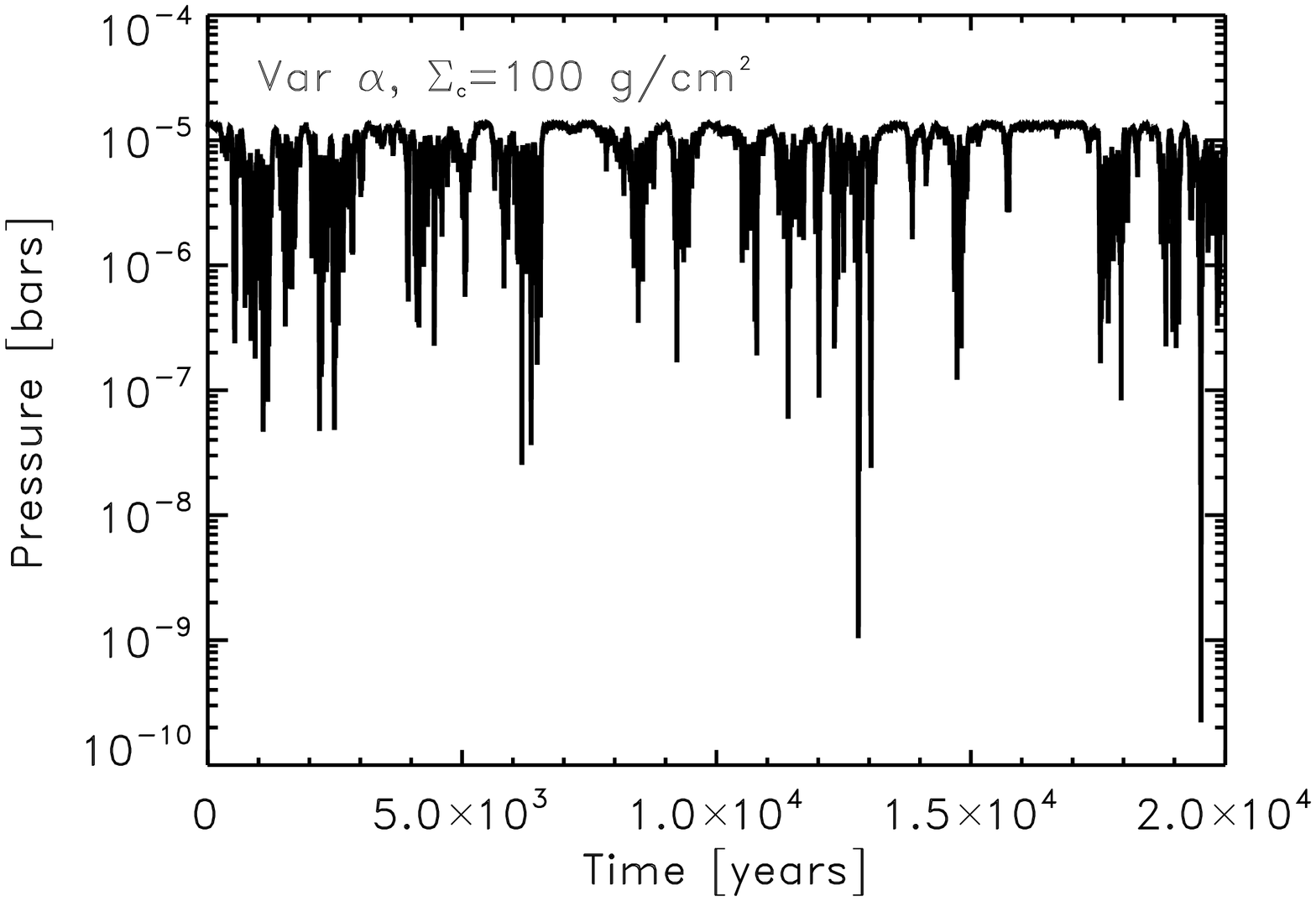}
\includegraphics[angle=90,width=3in]{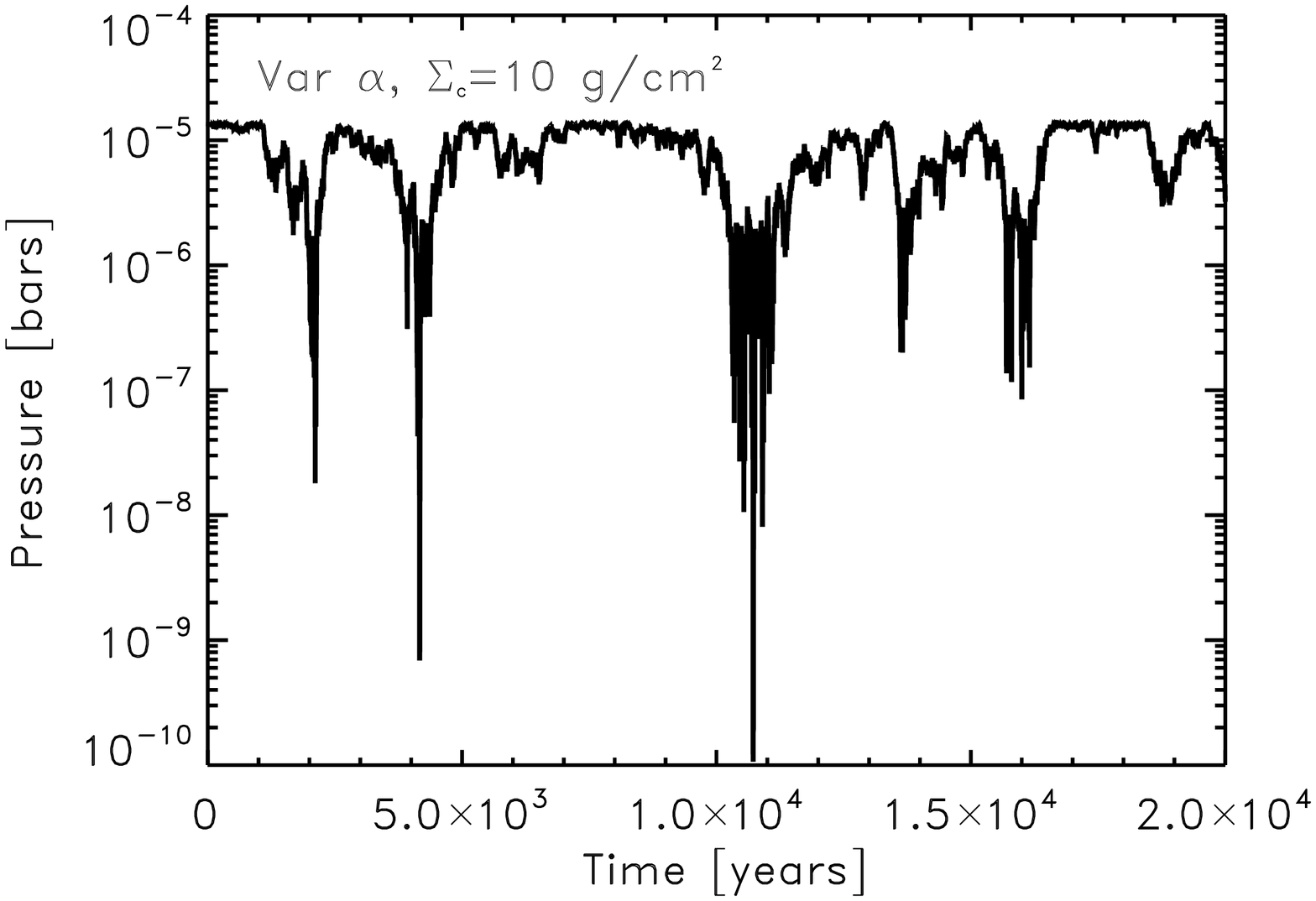}
\caption{The ambient pressures to which the particles in Figure 9 would have been exposed over the time periods shown.}
\end{figure}

\newpage
\begin{figure}
\includegraphics[angle=90,width=3in]{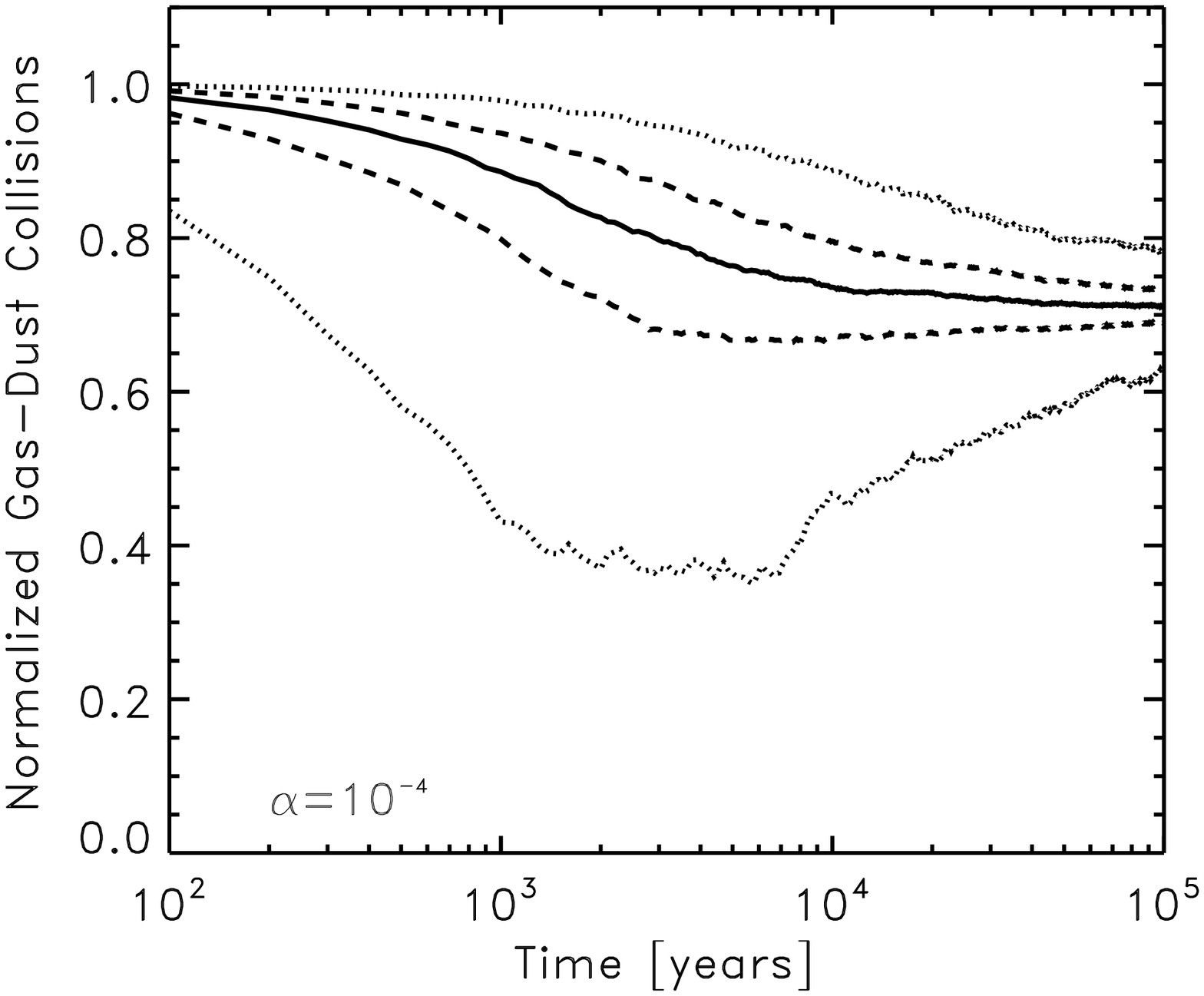}
\includegraphics[angle=90,width=3in]{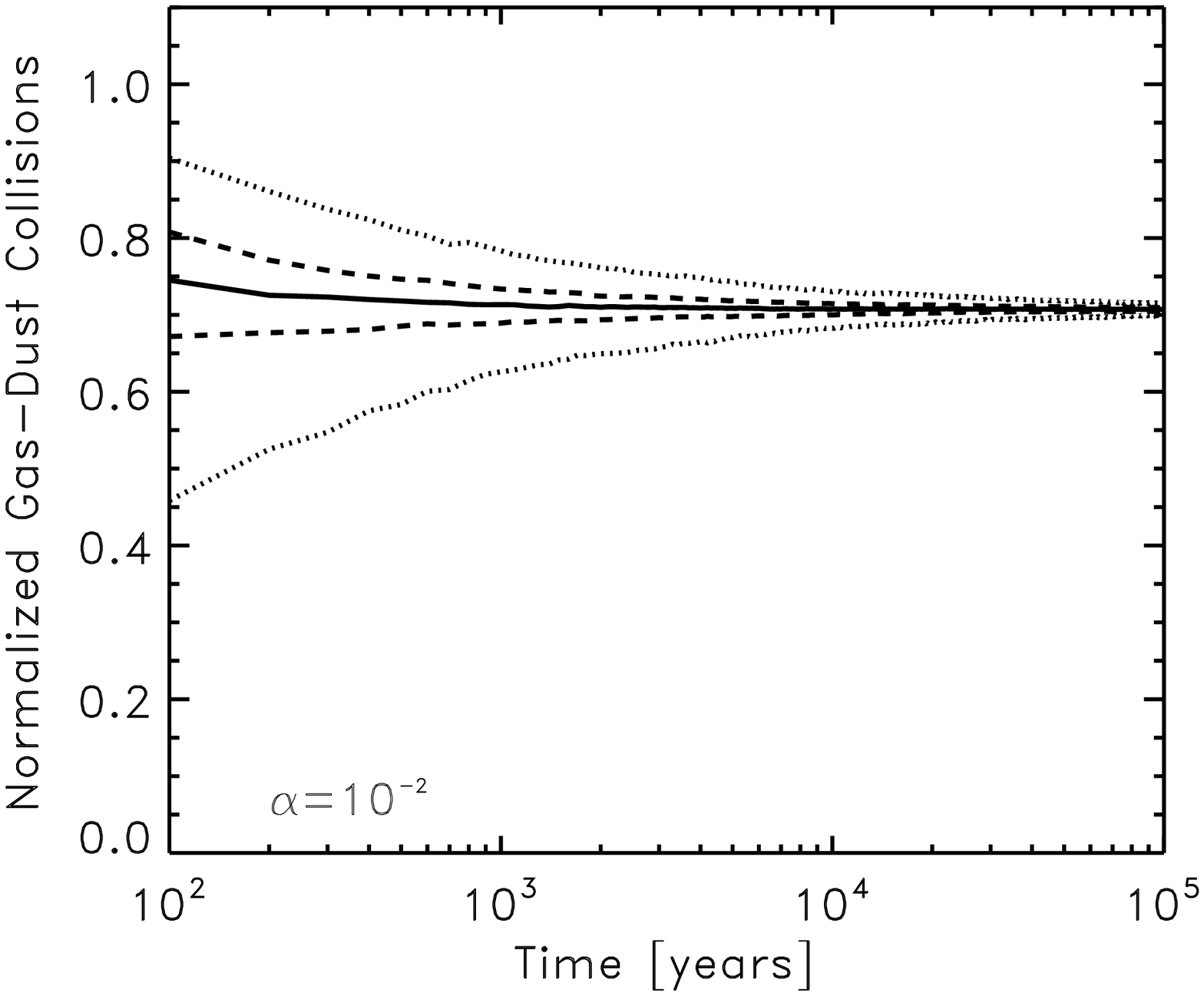}
\includegraphics[angle=90,width=3in]{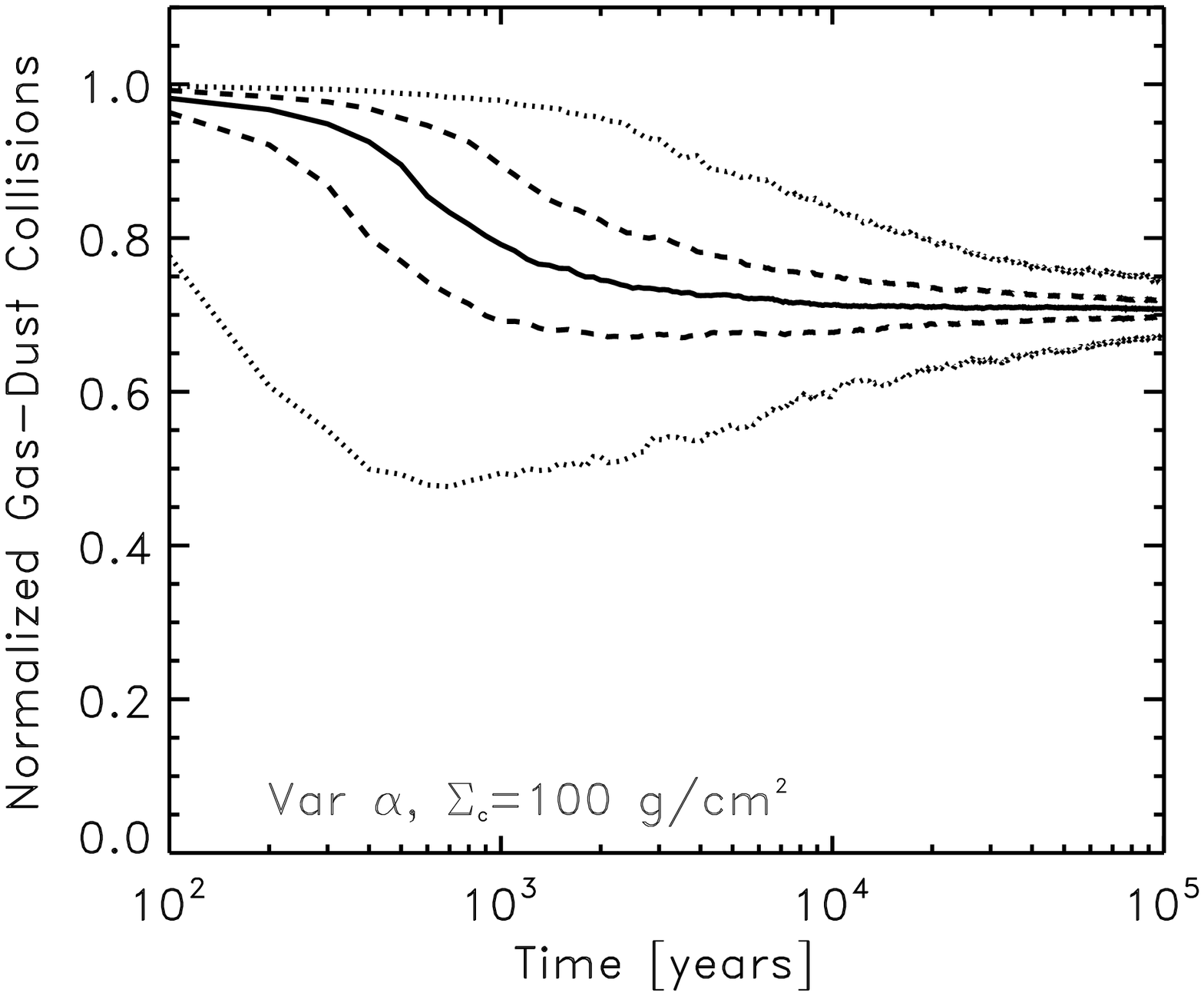}
\includegraphics[angle=90,width=3in]{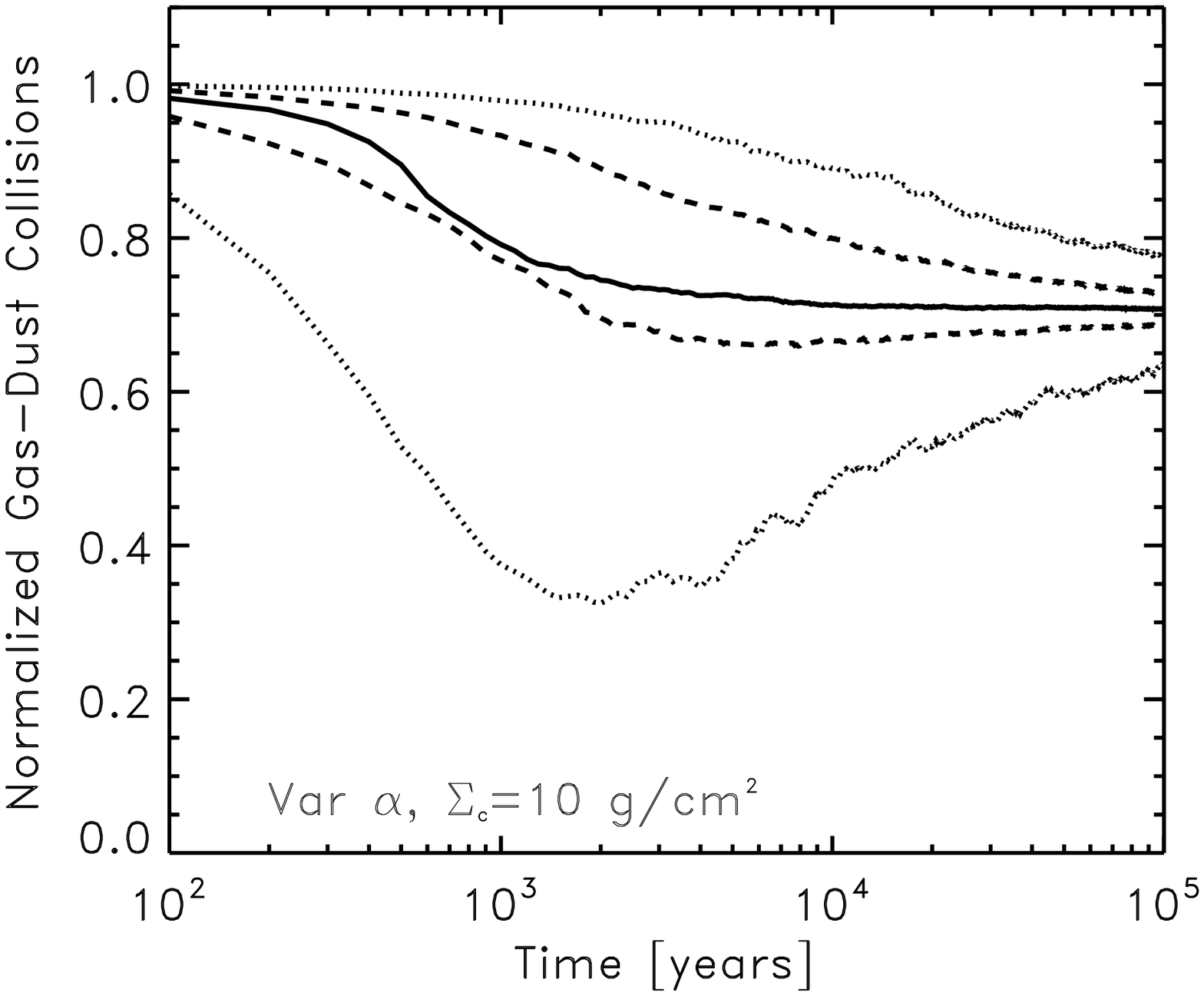}
\caption{Solid lines show temporal evolution of the number of collisions the average micron-sized dust particle would experience as it diffused in the vertical column of the disk over 10$^{5}$ years for the different $\alpha$ cases considered, normalized to the number of collisions it would have experienced if it remained at the disk midplane.  The bracketing dashed lines show the evolution for the 25\% and 75\% grains, while the dotted lines show those for the 1\% and 99\% grains.}
\end{figure}

\newpage
\begin{figure}
\includegraphics[angle=90,width=3in]{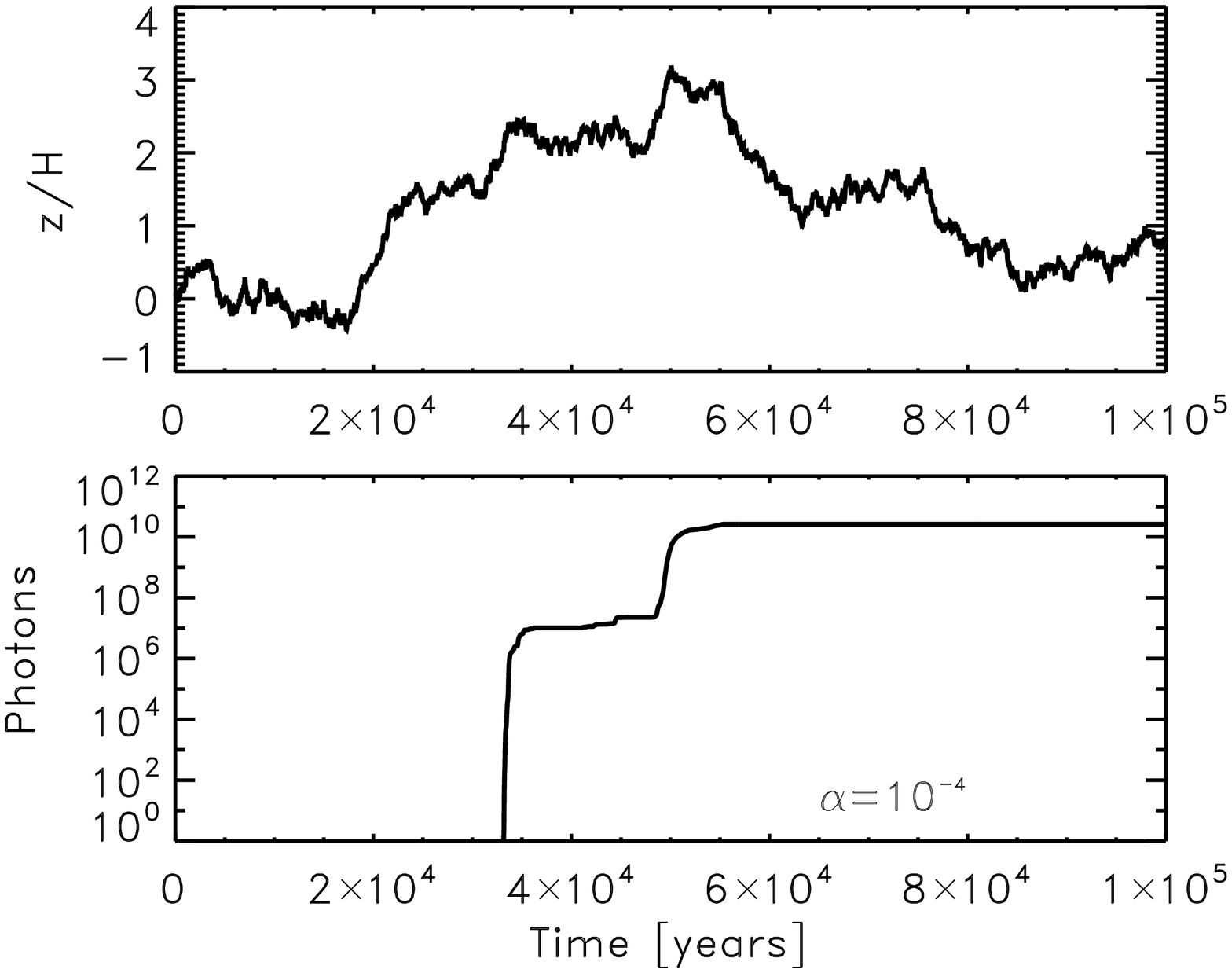}
\includegraphics[angle=90,width=3in]{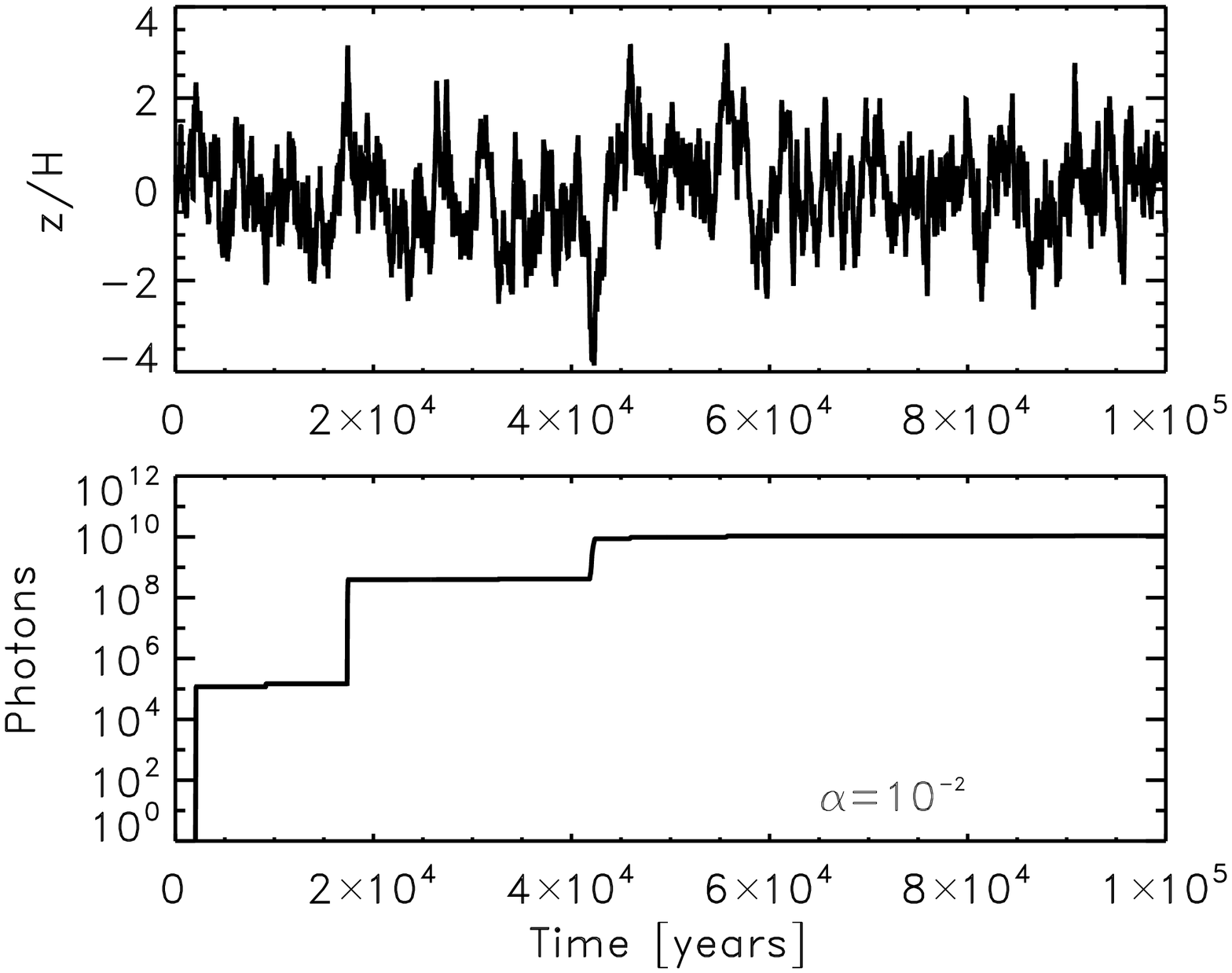}
\includegraphics[angle=90,width=3in]{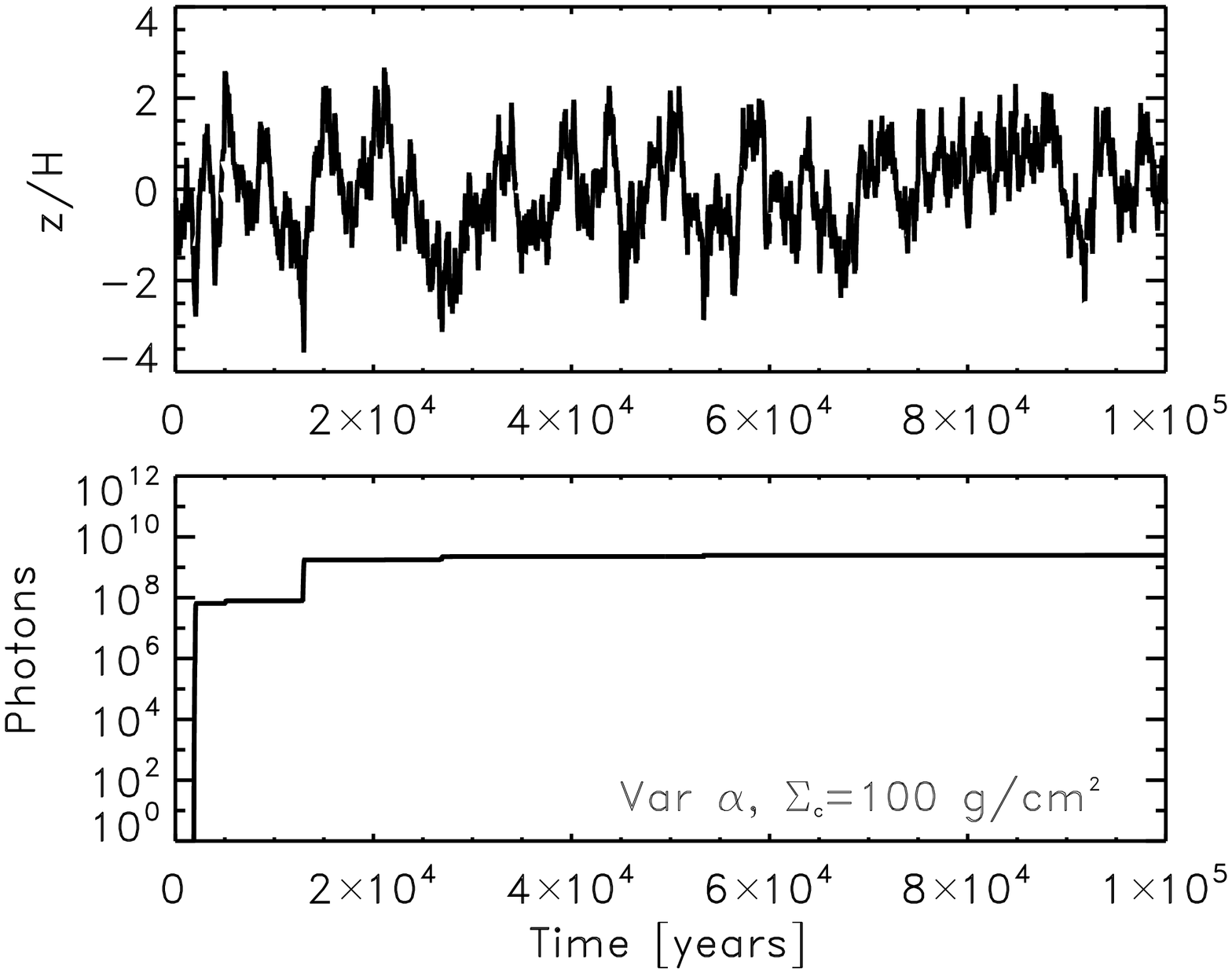}
\includegraphics[angle=90,width=3in]{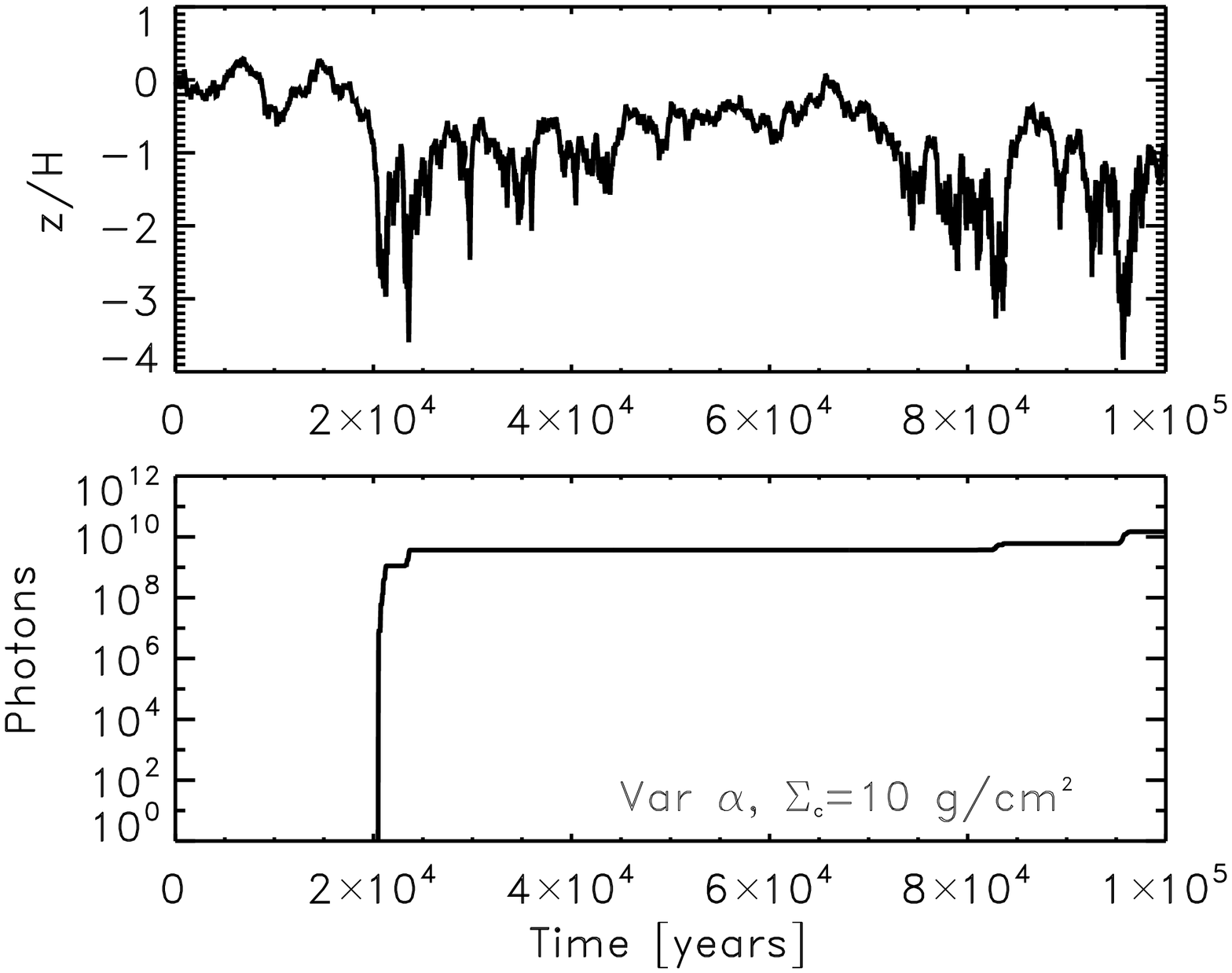}
\caption{Upper panels: Trajectories of individual micron-sized particles at 10 AU in a disk where $\Sigma$=200 g/cm$^{2}$ and $T$=88 K for each of the $\alpha$ cases considered here.  Lower panels: The cumulative number of UV photons incident on each grain over their lifetimes in the disk. An incident flux of UV photons of $G_{0}$=1 was assumed on the disk surface.}
\end{figure}

\newpage
\begin{figure}
\includegraphics[angle=90,width=3in]{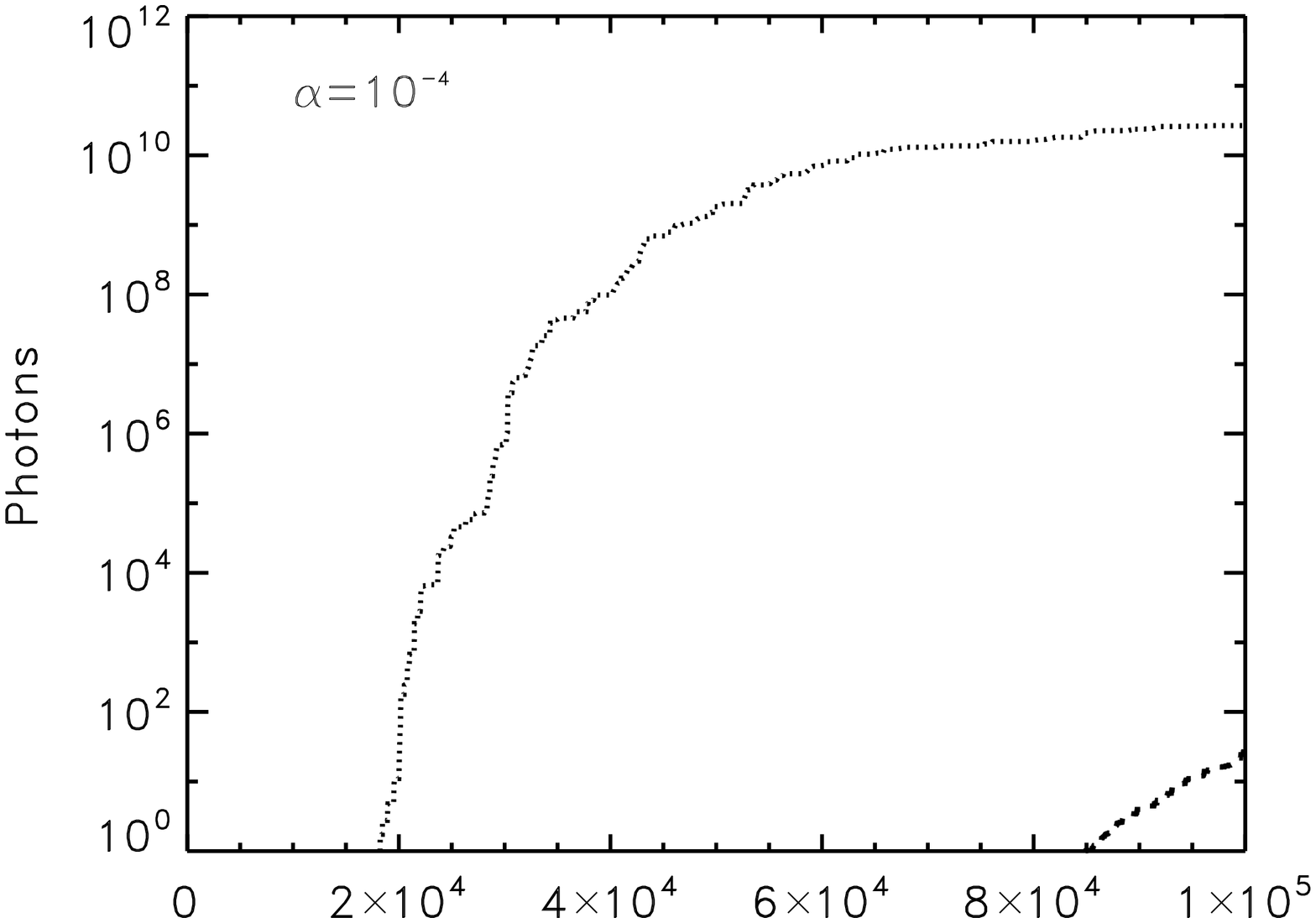}
\includegraphics[angle=90,width=3in]{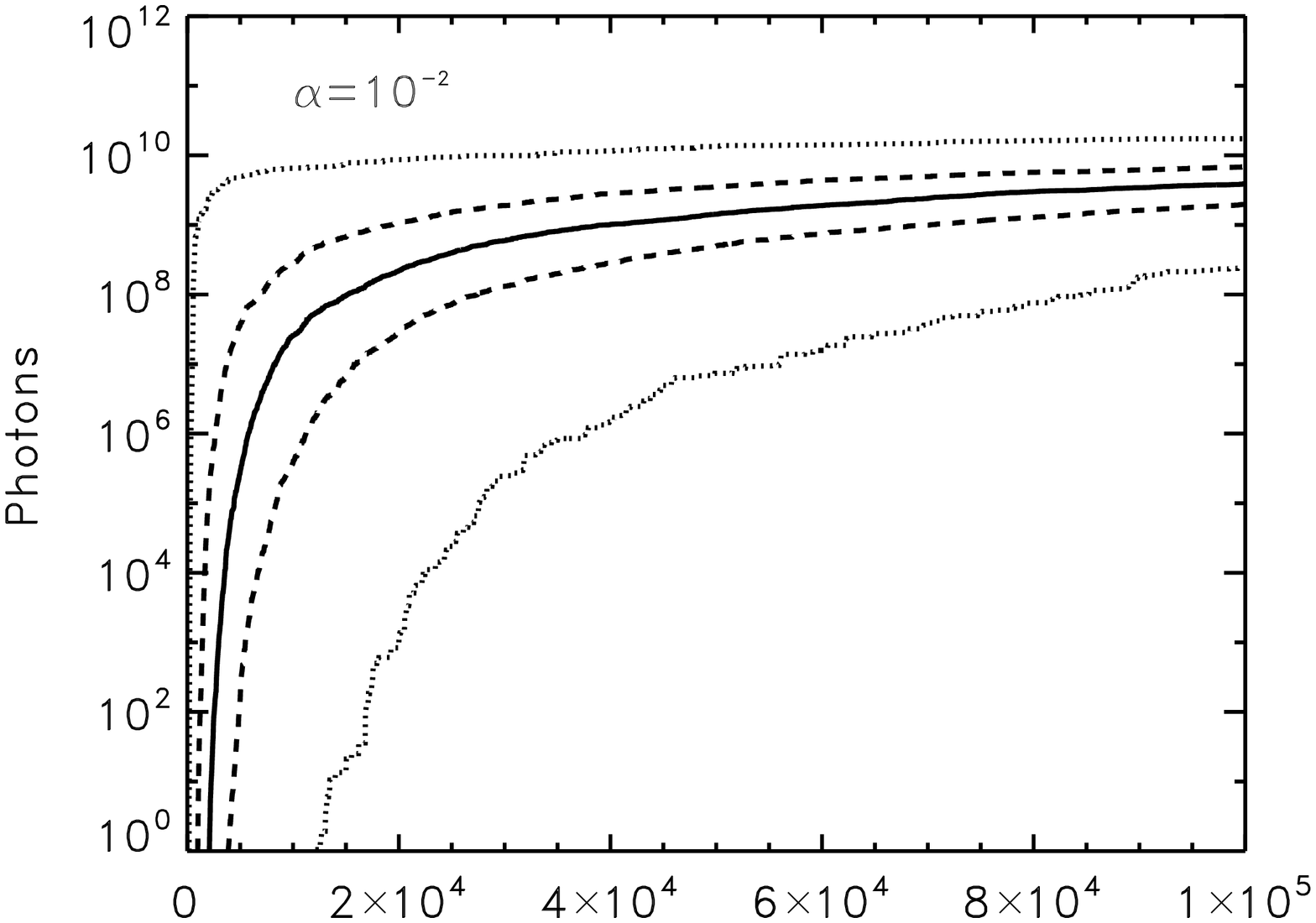}
\includegraphics[angle=90,width=3in]{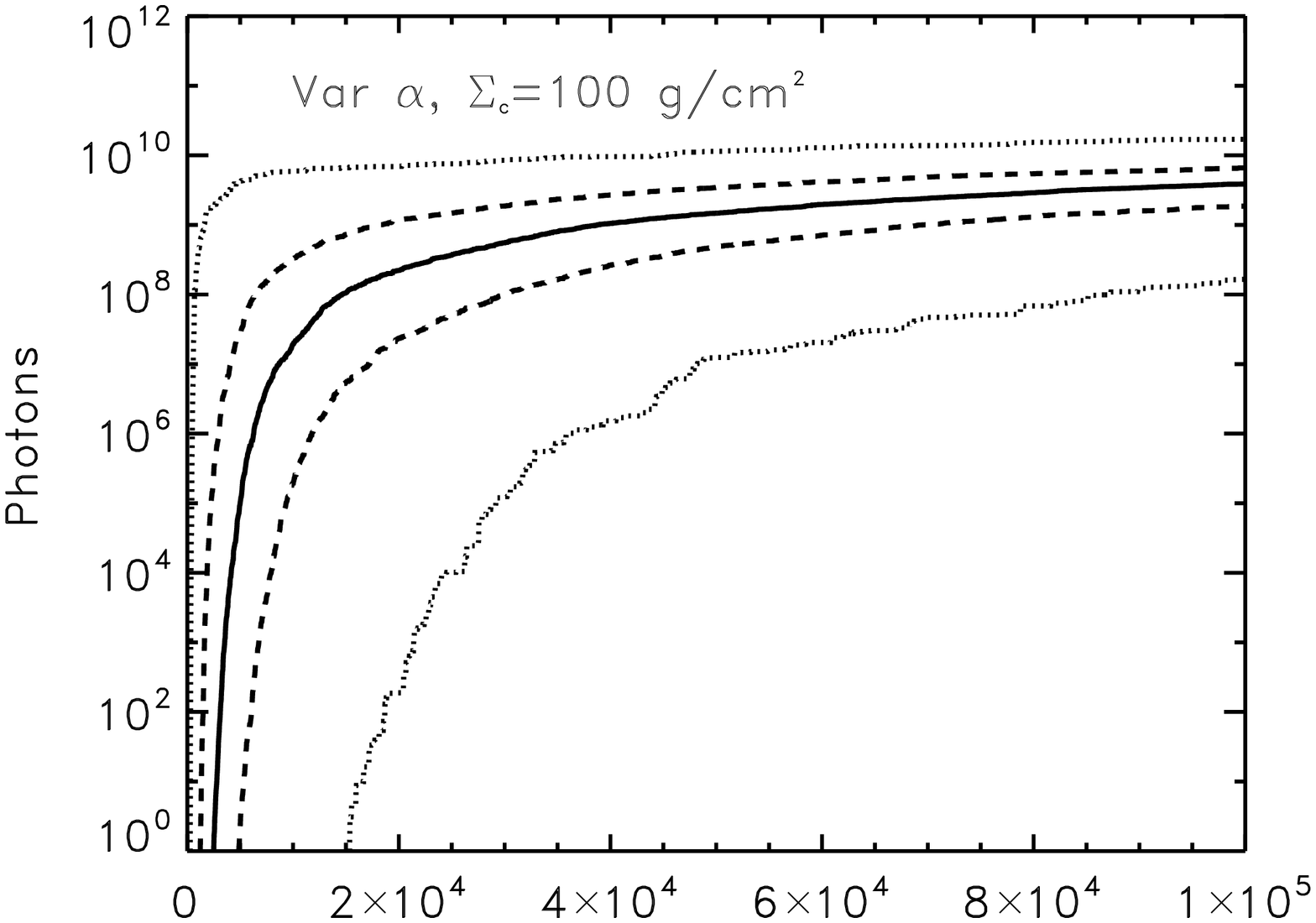}
\includegraphics[angle=90,width=3in]{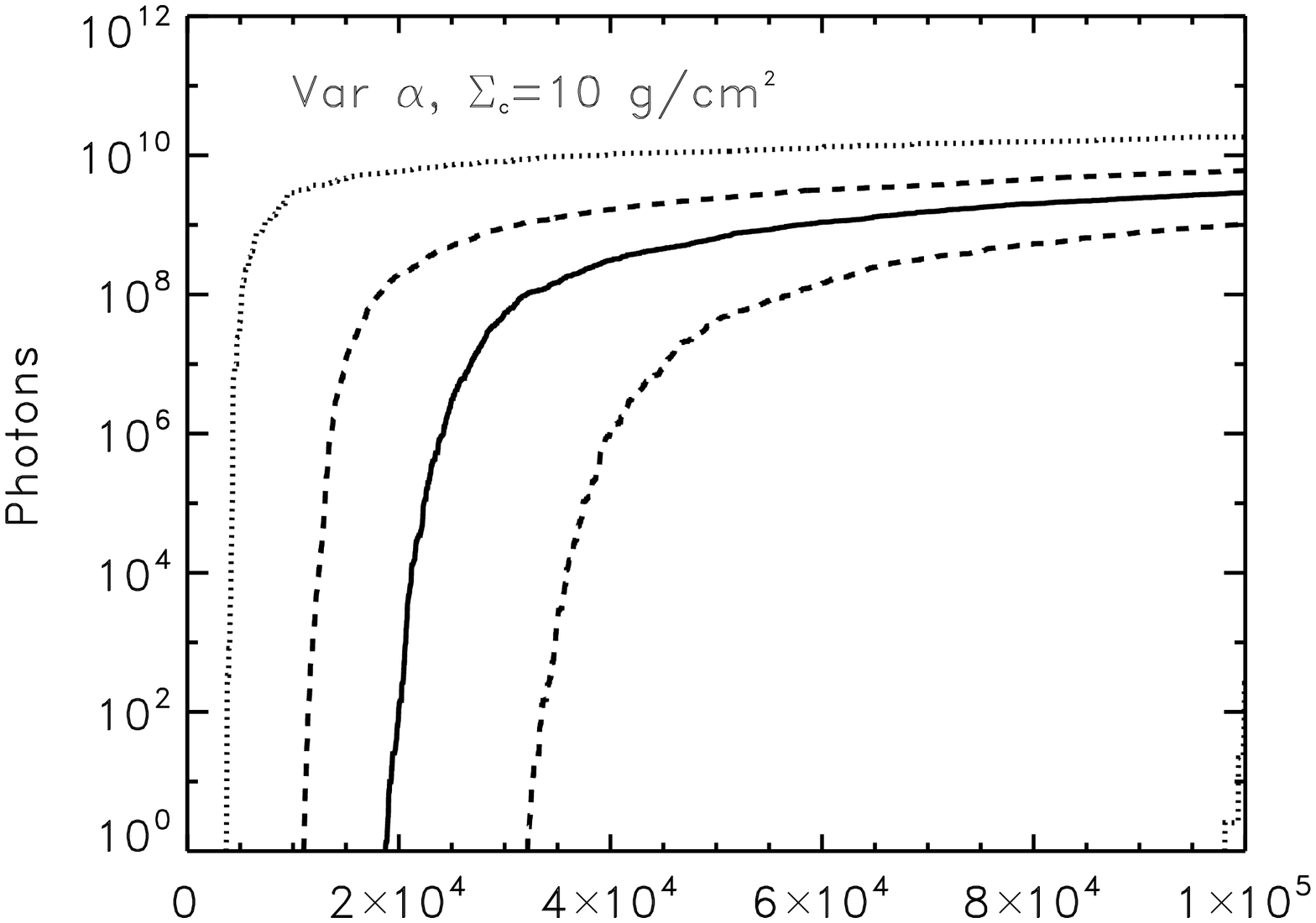}
\caption{Solid lines show temporal evolution of the number of incident photons on the median (50 \% rank) micron-sized dust particle as they vertically diffuse in the disk described in Figure 12.  The bracketing dashed lines show the cumulative number of photons for the 25\% and 75\% grains, while the dotted lines show those for the 1\% and 99\% grains. Note that $\alpha$=10$^{-4}$ plots show only the 75\% and 99\% grains while the 1\% grain can barely be seen in the lower right of the plot for the variable $\alpha$ with $\Sigma_{c}$=10 g/cm$^{2}$ case.}
\end{figure}

\end{document}